\documentclass[12pt]{article}
\begin{document}

\author{C.\ Bizdadea\thanks{%
e-mail address: bizdadea@central. ucv.ro}, E. M. Cioroianu\thanks{%
e-mail address: manache@central. ucv.ro}, A. C. Lungu and S. O. Saliu\thanks{%
e-mail addresses: osaliu@central.ucv.ro} \\
Faculty of Physics, University of Craiova\\
13 A. I.\ Cuza Str., Craiova 200585, Romania}
\title{No multi-graviton theories in the presence of a Dirac field}
\maketitle

\begin{abstract}
The cross-couplings among several massless spin-two fields (described in the
free limit by a sum of Pauli-Fierz actions) in the presence of a Dirac field
are investigated in the framework of the deformation theory based on local
BRST cohomology. Under the hypotheses of locality, smoothness of the
interactions in the coupling constant, Poincar\'{e} invariance, (background)
Lorentz invariance and the preservation of the number of derivatives on each
field, we prove that there are no consistent cross-interactions among
different gravitons in the presence of a Dirac field. The basic features of
the couplings between a single Pauli-Fierz field and a Dirac field are also
emphasized.

PACS number: 11.10.Ef
\end{abstract}

\section{Introduction}

Over the last twenty years there was a sustained effort for constructing
theories involving a multiplet of spin-two fields~\cite%
{cutwald1,wald2,ovrutwald,ancoann}. At the same time, various couplings of a
single massless spin-two field to other fields (including itself) have been
studied in~\cite%
{gupta54,kraich55,wein65,deser70,boul75,fangfr,berburgdam,wald86,hatffeyn,boulcqg}%
. In this context the impossibility of cross-interactions among several
Einstein gravitons under certain assumptions has recently been proved in~%
\cite{multi} by means of a cohomological approach based on the lagrangian
BRST symmetry~\cite{batvilk1,batvilk2,batvilk3,henproc,hencarte}. Moreover,
in~\cite{multi} the impossibility of cross-interactions among different
Einstein gravitons in the presence of a scalar field has also been shown.

The main aim of this paper is to investigate the cross-couplings among
several massless spin-two fields (described in the free limit by a sum of
Pauli-Fierz actions) in the presence of a Dirac field. More precisely, under
the hypotheses of locality, smoothness of the interactions in the coupling
constant, Poincar\'{e} invariance, (background) Lorentz invariance and the
preservation of the number of derivatives on each field, we prove that there
are no consistent cross-interactions among different gravitons in the
presence of a Dirac field. This result is obtained by using the deformation
technique~\cite{def} combined with the local BRST cohomology~\cite{gen1}. It
is well-known the fact that the spin-two field in metric formulation
(Einstein-Hilbert theory) cannot be coupled with a Dirac field. However, as
it will be seen, if we decompose the metric like $g_{\mu \nu }=\sigma _{\mu
\nu }+gh_{\mu \nu }$, where $\sigma _{\mu \nu }$ is the flat metric and $g$
is the coupling constant, then we can indeed couple Dirac spinors to $h_{\mu
\nu }$ in the space of formal series with the maximum derivative order equal
to one in $h_{\mu \nu }$, such that the final results agree with the usual
couplings between the spin-1/2 and the massless spin-two field in the
vierbein formulation~\cite{siegelfields}. Thus, our approach envisages two
different aspects. One is related to the couplings between the spin-two
fields and the Dirac field, while the other focuses on proving the
impossibility of cross-interactions among different gravitons via Dirac
spinors. In order to make the analysis as clear as possible, we initially
consider the case of the couplings between a single Pauli-Fierz field~\cite%
{pf} and a Dirac field. In this setting we compute the interaction terms to
order two in the coupling constant. Next, we prove the isomorphism between
the local BRST cohomologies corresponding to the Pauli-Fierz theory and
respectively to the linearized version of the vierbein formulation of the
spin-two field. Since the deformation procedure is controlled by the local
BRST cohomology of the free theory (in ghost number zero and one), the
previous isomorphism allows us to translate the results emerging from the
Pauli-Fierz formulation into the vierbein version and conversely. In this
manner we obtain that the first two orders of the interacting lagrangian
resulting from our setting originate in the development of the full
interacting lagrangian
\[
\mathcal{L}^{\left( \mathrm{int}\right) }=e\bar{\psi}\left( \mathrm{i}%
e_{a}^{\;\;\mu }\gamma ^{a}D_{\mu }\psi -m\psi \right) +egM\left( \bar{\psi}%
\psi \right) ,
\]
where $e_{a}^{\;\;\mu }$ are the vierbein fields, $e$ is the inverse of
their determinant, $e=\left( \det \left( e_{a}^{\;\;\mu }\right) \right)
^{-1}$, $D_{\mu }$ is the full covariant derivative and $M\left( \bar{\psi}%
\psi \right) $ is a polynomial in $\bar{\psi}\psi $. Here and in the sequel $%
g$ is the coupling constant (deformation parameter).The term $eM\left( \bar{%
\psi}\psi \right) $ is usually omitted in most of the textbooks on General
Relativity. However, it is consistent with the gauge symmetries of the
lagrangian $\mathcal{L}_{2}+\mathcal{L}^{\left( \mathrm{int}\right) }$,
where $\mathcal{L}_{2}$ is the full spin-two lagrangian in the vierbein
formulation. With this result at hand, we start from a finite sum of
Pauli-Fierz actions and a Dirac field, and prove that there are no
consistent cross-interactions between different gravitons in the presence of
a Dirac field.

This paper is organized in eight sections. In Section 2 we construct the
BRST symmetry of a free model with a single Pauli-Fierz field and a Dirac
field. Section 3 briefly addresses the deformation procedure based on BRST
symmetry. In Section 4 we compute the first two orders of the interactions
between one graviton and a Dirac spinor. Section 5 is dedicated to the proof
of the isomorphism between the local BRST cohomologies corresponding to the
Pauli-Fierz theory and respectively to the linearized version of the
vierbein formulation for the spin-two field. In Section 6 we connect the
results obtained in Section 4 to those from the vierbein formulation.
Section 7 is devoted to the proof of the fact that there are no consistent
cross-interactions among different gravitons in the presence of a Dirac
field. Section 8 exposes the main conclusions of the paper. The paper also
contains two appendix sections, in which some statements from the body of
the paper are proved.

\section{Free model: lagrangian formulation and BRST symmetry}

Our starting point is represented by a free model, whose lagrangian action
is written like the sum between the action of the linearized version of
Einstein-Hilbert gravity (the Pauli-Fierz action~\cite{pf}) and that of a
massive Dirac field
\begin{eqnarray}
S_{0}^{\mathrm{L}}\left[ h_{\mu \nu },\psi ,\bar{\psi}\right] &=&\int
d^{4}x\left( -\frac{1}{2}\left( \partial _{\mu }h_{\nu \rho }\right) \left(
\partial ^{\mu }h^{\nu \rho }\right) +\left( \partial _{\mu }h^{\mu \rho
}\right) \left( \partial ^{\nu }h_{\nu \rho }\right) \right.  \nonumber \\
&&\left. -\left( \partial _{\mu }h\right) \left( \partial _{\nu }h^{\nu \mu
}\right) +\frac{1}{2}\left( \partial _{\mu }h\right) \left( \partial ^{\mu
}h\right) +\bar{\psi}\left( \mathrm{i}\gamma ^{\mu }\left( \partial _{\mu
}\psi \right) -m\psi \right) \right)  \nonumber \\
&\equiv &\int d^{4}x\left( \mathcal{L}^{\left( \mathrm{PF}\right) }+\mathcal{%
L}_{0}^{(\mathrm{D})}\right) .  \label{fract}
\end{eqnarray}
Everywhere in the paper we use the flat Minkowski metric of `mostly plus'
signature, $\sigma _{\mu \nu }=\left( -+++\right) $. In the above $h$
denotes the trace of the Pauli-Fierz field, $h=\sigma _{\mu \nu }h^{\mu \nu
} $, and the fermionic fields $\psi $ and $\bar{\psi}$ are considered to be
complex (Dirac) spinors ($\bar{\psi}=\psi ^{\dagger }\gamma ^{0}$). We work
with the Dirac representation of the $\gamma $-matrices
\begin{equation}
\gamma _{\mu }^{\dagger }=\gamma _{0}\gamma _{\mu }\gamma _{0},\quad \mu =%
\overline{0,3},  \label{PFD4}
\end{equation}
where $\dagger $ signifies the operation of Hermitian conjugation. Action (%
\ref{fract}) possesses an irreducible and Abelian generating set of gauge
transformations
\begin{equation}
\delta _{\epsilon }h_{\mu \nu }=\partial _{(\mu }\epsilon _{\nu )},\quad
\delta _{\epsilon }\psi =\delta _{\epsilon }\bar{\psi}=0,  \label{PFD5}
\end{equation}
with $\epsilon _{\mu }$ bosonic gauge parameters. The parantheses signify
symmetrization; they are never divided by the number of terms: e.g., $%
\partial _{(\mu }\epsilon _{\nu )}=\partial _{\mu }\epsilon _{\nu }+\partial
_{\nu }\epsilon _{\mu }$, and the minimum number of terms is always used.
The same is valid with respect to the notation $\left[ \mu \cdots \nu \right]
$, which means antisymmetrization with respect to the indices between
brackets.

In order to construct the BRST symmetry for (\ref{fract}) we introduce the
fermionic ghosts $\eta _{\mu }$ corresponding to the gauge parameters $%
\epsilon _{\mu }$ and associate antifields with the original fields and
ghosts, respectively denoted by $\left\{ h^{*\mu \nu },\psi ^{*},\bar{\psi}%
^{*}\right\} $ and $\left\{ \eta ^{*\mu }\right\} $. (The statistics of the
antifields is opposite to that of the correlated fields/ghosts.) The
antifields of the Dirac fields are bosonic spinors, assumed to satisfy the
properties
\begin{equation}
\left( \bar{\psi}^{*}\right) ^{\dagger }\gamma ^{0}=-\psi ^{*},\quad \gamma
^{0}\left( \psi ^{*}\right) ^{\dagger }=-\bar{\psi}^{*}.  \label{PFD6}
\end{equation}
Since the gauge generators of the free theory under study are field
independent and irreducible, it follows that the BRST differential simply
decomposes into
\begin{equation}
s=\delta +\gamma ,  \label{PFD7}
\end{equation}
where $\delta $ represents the Koszul-Tate differential, graded by the
antighost number $\mathrm{agh}$ ($\mathrm{agh}\left( \delta \right) =-1$),
and $\gamma $ stands for the exterior derivative along the gauge orbits,
whose degree is named pure ghost number $\mathrm{pgh}$ ($\mathrm{pgh}\left(
\gamma \right) =1$). These two degrees do not interfere ($\mathrm{pgh}\left(
\delta \right) =0$, $\mathrm{agh}\left( \gamma \right) =0$). The overall
degree from the BRST complex is known as the ghost number $\mathrm{gh}$ and
is defined like the difference between the pure ghost number and the
antighost number, such that $\mathrm{gh}\left( \delta \right) =\mathrm{gh}%
\left( \gamma \right) =\mathrm{gh}\left( s\right) =1$. If we make the
notations
\begin{equation}
\Phi ^{\alpha _{0}}=\left( h_{\mu \nu },\psi ,\bar{\psi}\right) ,\;\Phi
_{\alpha _{0}}^{*}=\left( h^{*\mu \nu },\psi ^{*},\bar{\psi}^{*}\right) ,
\label{PFD9}
\end{equation}
then, according to the standard rules of the BRST formalism, the degrees of
the BRST generators are valued like
\begin{eqnarray}
\mathrm{agh}\left( \Phi ^{\alpha _{0}}\right) &=&\mathrm{agh}\left( \eta
_{\mu }\right) =0,\quad \mathrm{agh}\left( \Phi _{\alpha _{0}}^{*}\right)
=1,\quad \mathrm{agh}\left( \eta ^{*\mu }\right) =2,  \label{PFD10} \\
\mathrm{pgh}\left( \Phi ^{\alpha _{0}}\right) &=&0,\quad \mathrm{pgh}\left(
\eta _{\mu }\right) =1,\quad \mathrm{pgh}\left( \Phi _{\alpha
_{0}}^{*}\right) =\mathrm{pgh}\left( \eta ^{*\mu }\right) =0.  \label{PFD11}
\end{eqnarray}
The actions of the differentials $\delta $ and $\gamma $ on the generators
from the BRST complex are given by
\begin{eqnarray}
\delta h^{*\mu \nu } &=&2H^{\mu \nu },\;\delta \psi ^{*}=-\left( m\bar{\psi}+%
\mathrm{i}\partial _{\mu }\bar{\psi}\gamma ^{\mu }\right) ,  \label{PFD12} \\
\delta \bar{\psi}^{*} &=&-\left( \mathrm{i}\gamma ^{\mu }\partial _{\mu
}\psi -m\psi \right) ,\;\delta \eta ^{*\mu }=-2\partial _{\nu }h^{*\mu \nu },
\label{PFD13} \\
\delta \Phi ^{\alpha _{0}} &=&0=\delta \eta _{\mu },  \label{PFD14} \\
\gamma \Phi _{\alpha _{0}}^{*} &=&0=\gamma \eta ^{*\mu },  \label{PFD15} \\
\gamma h_{\mu \nu } &=&\partial _{(\mu }\eta _{\nu )},\quad \gamma \psi
=0=\gamma \bar{\psi},\quad \gamma \eta _{\mu }=0,  \label{PFD16}
\end{eqnarray}
where $H^{\mu \nu }$ is the linearized Einstein tensor
\begin{equation}
H^{\mu \nu }=K^{\mu \nu }-\frac{1}{2}\sigma ^{\mu \nu }K,  \label{PFD17a}
\end{equation}
with $K^{\mu \nu }$ and $K$ the linearized Ricci tensor and respectively the
linearized scalar curvature, both obtained from the linearized Riemann
tensor
\begin{eqnarray}
K_{\mu \nu \alpha \beta } &=&-\frac{1}{2}\left( \partial _{\mu }\partial
_{\alpha }h_{\nu \beta }+\partial _{\nu }\partial _{\beta }h_{\mu \alpha
}\right.  \nonumber \\
&&\left. -\partial _{\nu }\partial _{\alpha }h_{\mu \beta }-\partial _{\mu
}\partial _{\beta }h_{\nu \alpha }\right) ,  \label{PFD17b}
\end{eqnarray}
via its simple and double traces
\begin{equation}
K_{\mu \alpha }=\sigma ^{\nu \beta }K_{\mu \nu \alpha \beta },K=\sigma ^{\mu
\alpha }\sigma ^{\nu \beta }K_{\mu \nu \alpha \beta }.  \label{PFD17c}
\end{equation}

The BRST differential is known to have a canonical action in a structure
named antibracket and denoted by the symbol $\left( ,\right) $ ($s\cdot
=\left( \cdot ,\bar{S}\right) $), which is obtained by decreeing the
fields/ghosts respectively conjugated to the corresponding antifields. The
generator of the BRST symmetry is a bosonic functional of ghost number zero,
which is solution to the classical master equation $\left( \bar{S},\bar{S}%
\right) =0$. The full solution to the master equation for the free model
under study reads as
\begin{equation}
\bar{S}=S_{0}^{\mathrm{L}}\left[ h_{\mu \nu },\psi ,\bar{\psi}\right] +\int
d^{4}x\,h^{*\mu \nu }\partial _{(\mu }\eta _{\nu )}.  \label{PFD18}
\end{equation}

\section{Deformation of the solution to the master equation: a brief review}

We begin with a ``free'' gauge theory, described by a lagrangian action $%
S_{0}^{\mathrm{L}}\left[ \Phi ^{\alpha _{0}}\right] $, invariant under some
gauge transformations $\delta _{\epsilon }\Phi ^{\alpha _{0}}=Z_{\;\;\alpha
_{1}}^{\alpha _{0}}\epsilon ^{\alpha _{1}}$, i.e. $\frac{\delta S_{0}^{%
\mathrm{L}}}{\delta \Phi ^{\alpha _{0}}}Z_{\;\;\alpha _{1}}^{\alpha _{0}}=0$%
, and consider the problem of constructing consistent interactions among the
fields $\Phi ^{\alpha _{0}}$ such that the couplings preserve both the field
spectrum and the original number of gauge symmetries. This matter is
addressed by means of reformulating the problem of constructing consistent
interactions as a deformation problem of the solution to the master equation
corresponding to the ``free'' theory~\cite{def}. Such a reformulation is
possible due to the fact that the solution to the master equation contains
all the information on the gauge structure of the theory. If an interacting
gauge theory can be consistently constructed, then the solution $\bar{S}$ to
the master equation $\left( \bar{S},\bar{S}\right) =0$ associated with the
``free'' theory can be deformed into a solution $S$
\begin{eqnarray}
\bar{S}\rightarrow S &=&\bar{S}+gS_{1}+g^{2}S_{2}+\cdots  \nonumber \\
&=&\bar{S}+g\int d^{D}x\,a+g^{2}\int d^{D}x\,b+\cdots ,  \label{PFD2.2}
\end{eqnarray}
of the master equation for the deformed theory
\begin{equation}
\left( S,S\right) =0,  \label{PFD2.3}
\end{equation}
such that both the ghost and antifield spectra of the initial theory are
preserved. The equation (\ref{PFD2.3}) splits, according to the various
orders in the coupling constant (deformation parameter) $g$, into a tower of
equations:
\begin{eqnarray}
\left( \bar{S},\bar{S}\right) &=&0,  \label{PFD2.4} \\
2\left( S_{1},\bar{S}\right) &=&0,  \label{PFD2.5} \\
2\left( S_{2},\bar{S}\right) +\left( S_{1},S_{1}\right) &=&0,  \label{PFD2.6}
\\
\left( S_{3},\bar{S}\right) +\left( S_{1},S_{2}\right) &=&0,  \label{PFD2.7}
\\
&&\vdots  \nonumber
\end{eqnarray}

The equation (\ref{PFD2.4}) is fulfilled by hypothesis. The next one
requires that the first-order deformation of the solution to the master
equation, $S_{1}$, is a cocycle of the ``free'' BRST differential $s\cdot
=\left( \cdot ,\bar{S}\right) $. However, only cohomologically non-trivial
solutions to (\ref{PFD2.5}) should be taken into account, as the BRST-exact
ones can be eliminated by some (in general non-linear) field redefinitions.
This means that $S_{1}$ pertains to the ghost number zero cohomological
space of $s$, $H^{0}\left( s\right) $, which is generically non-empty due to
its isomorphism to the space of physical observables of the ``free'' theory.
It has been shown (on behalf of the triviality of the antibracket map in the
cohomology of the BRST differential) that there are no obstructions in
finding solutions to the remaining equations, namely (\ref{PFD2.6}) and (\ref%
{PFD2.7}) and so on. However, the resulting interactions may be non-local,
and there might even appear obstructions if one insists on their locality.
The analysis of these obstructions can be done by means of standard
cohomological techniques.

\section{Consistent interactions between the spin-two field and the massive
Dirac field}

\subsection{Standard material: $H\left( \protect\gamma \right) $ and $%
H\left( \protect\delta |d\right) $}

This section is devoted to the investigation of consistent cross-couplings
that can be introduced between a spin-two field and a massive Dirac field.
This matter is addressed in the context of the antifield-BRST deformation
procedure briefly addressed in the above and relies on computing the
solutions to the equations (\ref{PFD2.5})--(\ref{PFD2.7}), etc., with the
help of the free BRST cohomology.

For obvious reasons, we consider only smooth, local, (background) Lorentz
invariant and, moreover, Poincar\'{e} invariant quantities (i.e. we do not
allow explicit dependence on the spacetime coordinates). The smoothness of
the deformations refers to the fact that the deformed solution to the master
equation (\ref{PFD2.2}) is smooth in the coupling constant $g$ and reduces
to the original solution (\ref{PFD18}) in the free limit $g=0$. In addition,
we require the conservation of the number of derivatives on each field (this
condition is frequently met in the literature; for instance, see the case of
cross-interactions for a collection of Pauli-Fierz fields~\cite{multi} or
the couplings between the Pauli-Fierz and the massless Rarita-Schwinger
fields~\cite{boulcqg}). If we make the notation $S_{1}=\int d^{4}x\,a$, with
$a$ a local function, then the equation (\ref{PFD2.5}), which we have seen
that controls the first-order deformation, takes the local form
\begin{equation}
sa=\partial _{\mu }m^{\mu },\quad \mathrm{gh}\left( a\right) =0,\quad
\varepsilon \left( a\right) =0,  \label{PFD3.1}
\end{equation}
for some local $m^{\mu }$ and it shows that the non-integrated density of
the first-order deformation pertains to the local cohomology of the BRST
differential in ghost number zero, $a\in H^{0}\left( s|d\right) $, where $d$
denotes the exterior spacetime differential. The solution to the equation (%
\ref{PFD3.1}) is unique up to $s$-exact pieces plus divergences
\begin{equation}
a\rightarrow a+sb+\partial _{\mu }n^{\mu },\;\mathrm{gh}\left( b\right)
=-1,\;\varepsilon \left( b\right) =1,\;\mathrm{gh}\left( n^{\mu }\right)
=0,\;\varepsilon \left( n^{\mu }\right) =0.  \label{PFD3.1a}
\end{equation}
At the same time, if the general solution of (\ref{PFD3.1}) is found to be
completely trivial, $a=sb+\partial _{\mu }n^{\mu }$, then it can be made to
vanish $a=0$.

In order to analyze the equation (\ref{PFD3.1}), we develop $a$ according to
the antighost number
\begin{equation}
a=\sum\limits_{i=0}^{I}a_{i},\quad \mathrm{agh}\left( a_{i}\right) =i,\quad
\mathrm{gh}\left( a_{i}\right) =0,\quad \varepsilon \left( a_{i}\right) =0,
\label{PFD3.2}
\end{equation}%
and take this decomposition to stop at some finite value $I$ of the
antighost number. The fact that $I$ in (\ref{PFD3.2}) is finite can be
argued like in~\cite{multi}. Inserting the above expansion into the equation
(\ref{PFD3.1}) and projecting it on the various values of the antighost
number with the help of the splitting (\ref{PFD7}), we obtain the tower of
equations
\begin{eqnarray}
\gamma a_{I} &=&\partial _{\mu }\stackrel{\left( I\right) }{m}^{\mu
},
\label{PFD3.3} \\
\delta a_{I}+\gamma a_{I-1} &=&\partial _{\mu }\stackrel{\left( I-1\right) }{m%
}^{\mu },  \label{PFD3.4} \\
\delta a_{i}+\gamma a_{i-1} &=&\partial _{\mu }\stackrel{\left( i-1\right) }{m%
}^{\mu },\quad 1\leq i\leq I-1,  \label{PFD3.5}
\end{eqnarray}%
where $\left( \stackrel{\left( i\right) }{m}^{\mu }\right)
_{i=\overline{0,I}}
$ are some local currents with $\mathrm{agh}\left( \stackrel{\left( i\right) }%
{m}^{\mu }\right) =i$. Moreover, according to the general result from~\cite%
{multi} in the absence of the collection indices, the equation (\ref{PFD3.3}%
) can be replaced\footnote{%
This is because the presence of the matter fields does not modify the
general results on $H\left( \gamma \right) $ presented in~\cite{multi}.} in
strictly positive antighost numbers by
\begin{equation}
\gamma a_{I}=0,\quad I>0.  \label{PFD3.6}
\end{equation}%
Due to the second-order nilpotency of $\gamma $ ($\gamma ^{2}=0$), the
solution to the equation (\ref{PFD3.6}) is clearly unique up to $\gamma $%
-exact contributions
\begin{equation}
a_{I}\rightarrow a_{I}+\gamma b_{I},\quad \mathrm{agh}\left( b_{I}\right)
=I,\quad \mathrm{pgh}\left( b_{I}\right) =I-1,\quad \varepsilon \left(
b_{I}\right) =1.  \label{PFDr68}
\end{equation}%
Meanwhile, if it turns out that $a_{I}$ reduces to $\gamma $-exact terms
only, $a_{I}=\gamma b_{I}$, then it can be made to vanish, $a_{I}=0$. The
non-triviality of the first-order deformation $a$ is thus translated at its
highest antighost number component into the requirement that $a_{I}\in
H^{I}\left( \gamma \right) $, where $H^{I}\left( \gamma \right) $ denotes
the cohomology of the exterior longitudinal derivative $\gamma $ in pure
ghost number equal to $I$. So, in order to solve the equation (\ref{PFD3.1})
(equivalent with (\ref{PFD3.6}) and (\ref{PFD3.4}) and (\ref{PFD3.5})), we
need to compute the cohomology of $\gamma $, $H\left( \gamma \right) $, and,
as it will be made clear below, also the local cohomology of $\delta $ in
pure ghost number zero, $H\left( \delta |d\right) $.

Using the results on the cohomology of the exterior longitudinal
differential for a collection of Pauli-Fierz fields~\cite{multi}, as well as
the definitions (\ref{PFD15}) and (\ref{PFD16}), we can state that $H\left(
\gamma \right) $ is generated on the one hand by $\Phi _{\alpha _{0}}^{*}$, $%
\eta _{\mu }^{*}$, $\psi $, $\bar{\psi}$ and $K_{\mu \nu \alpha \beta }$
together with all of their spacetime derivatives and, on the other hand, by
the ghosts $\eta _{\mu }$ and $\partial _{[\mu }\eta _{\nu ]}$. So, the most
general (and non-trivial), local solution to (\ref{PFD3.6}) can be written,
up to $\gamma $-exact contributions, as
\begin{equation}
a_{I}=\alpha _{I}\left( \left[ \psi \right] ,\left[ \bar{\psi}\right] ,\left[
K_{\mu \nu \alpha \beta }\right] ,\left[ \Phi _{\alpha _{0}}^{*}\right] ,%
\left[ \eta _{\mu }^{*}\right] \right) \omega ^{I}\left( \eta _{\mu
},\partial _{[\mu }\eta _{\nu ]}\right) ,  \label{PFD3.10}
\end{equation}
where the notation $f\left( \left[ q\right] \right) $ means that $f$ depends
on $q$ and its derivatives up to a finite order, while $\omega ^{I}$ denotes
the elements of a basis in the space of polynomials with pure ghost number $%
I $ in the corresponding ghosts and their antisymmetrized first-order
derivatives. The objects $\alpha _{I}$ have the pure ghost number equal to
zero and are required to fulfill the property $\mathrm{agh}\left( \alpha
_{I}\right) =I$ in order to ensure that the ghost number of $a_{I}$ is equal
to zero. Since they have a bounded number of derivatives and a finite
antighost number, $\alpha _{I}$ are actually polynomials in the linearized
Riemann tensor, in the antifields, in all of their derivatives, as well as
in the derivatives of the Dirac fields. The anticommuting behaviour of the
Dirac spinors induces that $\alpha _{I}$ are polynomials also in the
undifferentiated Dirac fields, so we conclude that these elements exhibit a
polynomial character in all of their arguments. Due to their $\gamma $%
-closeness, $\gamma \alpha _{I}=0$, $\alpha _{I}$ will be called ``invariant
polynomials''. In zero antighost number, the invariant polynomials are
polynomials in the linearized Riemann tensor $K_{\mu \nu \alpha \beta }$, in
the Dirac spinors, as well as in their derivatives.

Inserting (\ref{PFD3.10}) in (\ref{PFD3.4}) we obtain that a necessary (but
not sufficient) condition for the existence of (non-trivial) solutions $%
a_{I-1}$ is that the invariant polynomials $\alpha _{I}$ are (non-trivial)
objects from the local cohomology of the Koszul-Tate differential $H\left(
\delta |d\right) $ in pure ghost number zero and in strictly positive
antighost numbers $I>0$%
\begin{equation}
\delta \alpha _{I}=\partial _{\mu }\stackrel{\left( I-1\right)
}{j}^{\mu },\quad \mathrm{agh}\left( \stackrel{\left( I-1\right)
}{j}^{\mu }\right) =I-1,\quad \mathrm{pgh}\left( \stackrel{\left(
I-1\right) }{j}^{\mu }\right) =0.  \label{PFD3.11}
\end{equation}
We recall that $H\left( \delta |d\right) $ is completely trivial in both
strictly positive antighost \textit{and} pure ghost numbers (for instance,
see~\cite{gen1}, Theorem 5.4 and~\cite{commun1}). Using the fact that the
Cauchy order of the free theory under study is equal to two together with
the general results from~\cite{gen1}, according to which the local
cohomology of the Koszul-Tate differential in pure ghost number zero is
trivial in antighost numbers strictly greater than its Cauchy order, we can
state that
\begin{equation}
H_{J}\left( \delta |d\right) =0\quad \mathrm{for\;all\;}J>2,  \label{PFD3.12}
\end{equation}
where $H_{J}\left( \delta |d\right) $ represents the local cohomology of the
Koszul-Tate differential in zero pure ghost number and in antighost number $%
J $. An interesting property of invariant polynomials for the free model
under study is that if an invariant polynomial $\alpha _{J}$, with $\mathrm{%
agh}\left( \alpha _{J}\right) =J\geq 2$, is trivial in $H_{J}\left( \delta
|d\right) $, then it can be taken to be trivial also in $H_{J}^{\mathrm{inv}%
}\left( \delta |d\right) $, i.e.
\begin{equation}
\left( \alpha _{J}=\delta b_{J+1}+\partial _{\mu }\stackrel{(J)}{c}^{\mu },\;%
\mathrm{agh}\left( \alpha _{J}\right) =J\geq 2\right) \Rightarrow
\alpha _{J}=\delta \beta _{J+1}+\partial _{\mu
}\stackrel{(J)}{\gamma }^{\mu }, \label{PFD3.12ax}
\end{equation}
with both $\beta _{J+1}$ and $\stackrel{(J)}{\gamma }^{\mu }$
invariant polynomials. Here, $H_{J}^{\mathrm{inv}}\left( \delta
|d\right) $ denotes the invariant characteristic cohomology (the
local cohomology of the Koszul-Tate differential in the space of
invariant polynomials) in antighost number $J$. This property is
proved in~\cite{multi} in the case of a collection of Pauli-Fierz
fields and remains valid in the case considered here since the
matter fields do not carry gauge symmetries, so we can write that
\begin{equation}
H_{J}^{\mathrm{inv}}\left( \delta |d\right) =0\quad \mathrm{for\;all\;}J>2.
\label{PFD3.12x}
\end{equation}
For the same reason the antifields of the matter fields can bring only
trivial contributions to $H_{J}\left( \delta |d\right) $ and $H_{J}^{\mathrm{%
inv}}\left( \delta |d\right) $ for $J\geq 2$, so the results from~\cite%
{multi} concerning both $H_{2}\left( \delta |d\right) $ in pure ghost number
zero and $H_{2}^{\mathrm{inv}}\left( \delta |d\right) $ remain valid. These
cohomological spaces are still spanned by the undifferentiated antifields
corresponding to the ghosts
\begin{equation}
H_{2}\left( \delta |d\right) \;\mathrm{and}\;H_{2}^{\mathrm{inv}}\left(
\delta |d\right) :\left( \eta ^{*\mu }\right) .  \label{PFD3.12b}
\end{equation}
In contrast to the groups $\left( H_{J}\left( \delta |d\right) \right)
_{J\geq 2}$ and $\left( H_{J}^{\mathrm{inv}}\left( \delta |d\right) \right)
_{J\geq 2}$, which are finite-dimensional, the cohomology $H_{1}\left(
\delta |d\right) $ in pure ghost number zero, known to be related to global
symmetries and ordinary conservation laws, is infinite-dimensional since the
theory is free. Moreover, $H_{1}\left( \delta |d\right) $ involves
non-trivially the antifields of the matter fields.

The previous results on $H\left( \delta |d\right) $ and $H^{\mathrm{inv}%
}\left( \delta |d\right) $ in strictly positive antighost numbers are
important because they control the obstructions to removing the antifields
from the first-order deformation. More precisely, based on the formulas (\ref%
{PFD3.11})--(\ref{PFD3.12x}), one can successively eliminate all the pieces
of antighost number strictly greater that two from the non-integrated
density of the first-order deformation by adding only trivial terms, so one
can take, without loss of non-trivial objects, the condition $I\leq 2$ in
the decomposition (\ref{PFD3.2}). The proof of this statement can be
realized like in~\cite{multi}. In addition, the last representative is of
the form (\ref{PFD3.10}), where the invariant polynomial is necessarily a
non-trivial object from $H_{2}^{\mathrm{inv}}\left( \delta |d\right) $ for $%
I=2$, and respectively from $H_{1}\left( \delta |d\right) $ for $I=1$.

\subsection{First-order deformation}

In the case $I=2$ the non-integrated density of the first-order deformation (%
\ref{PFD3.2}) becomes
\begin{equation}
a=a_{0}+a_{1}+a_{2}.  \label{PFD3.12t}
\end{equation}
We can further decompose $a$ in a natural manner as a sum between three
kinds of deformations
\begin{equation}
a=a^{\left( \mathrm{PF}\right) }+a^{\left( \mathrm{int}\right) }+a^{\left(
\mathrm{Dirac}\right) },  \label{PFD3.12a}
\end{equation}
where $a^{\left( \mathrm{PF}\right) }$ contains only
fields/ghosts/antifields from the Pauli-Fierz sector, $a^{\left( \mathrm{int}%
\right) }$ describes the cross-interactions between the two theories (so it
effectively mixes both sectors), and $a^{\left( \mathrm{Dirac}\right) }$
involves only the Dirac sector. The component $a^{\left( \mathrm{PF}\right)
} $ is completely known (for a detailed analysis see~\cite{multi}) and
satisfies individually an equation of the type (\ref{PFD3.1}). It admits a
decomposition similar to (\ref{PFD3.12t})
\begin{equation}
a^{\left( \mathrm{PF}\right) }=a_{0}^{\left( \mathrm{PF}\right)
}+a_{1}^{\left( \mathrm{PF}\right) }+a_{2}^{\left( \mathrm{PF}\right) },
\label{descPF}
\end{equation}
where
\begin{eqnarray}
a_{2}^{\left( \mathrm{PF}\right) } &=&\frac{1}{2}\eta ^{*\mu }\eta ^{\nu
}\partial _{\left[ \mu \right. }\eta _{\left. \nu \right] },  \label{PFa2} \\
a_{1}^{\left( \mathrm{PF}\right) } &=&h^{*\mu \rho }\left( \left( \partial
_{\rho }\eta ^{\nu }\right) h_{\mu \nu }-\eta ^{\nu }\partial _{[\mu }h_{\nu
]\rho }\right) ,  \label{PFa1}
\end{eqnarray}
and $a_{0}^{\left( \mathrm{PF}\right) }$ is the cubic vertex of the
Einstein-Hilbert lagrangian plus a cosmological term\footnote{%
The terms $a_{2}^{\left( \mathrm{PF}\right) }$ and $a_{1}^{\left( \mathrm{PF}%
\right) }$ given in (\ref{PFa2}) and (\ref{PFa1}) differ from the
corresponding ones in~\cite{multi} by a $\gamma $-exact and respectively a $%
\delta $-exact contribution. However, the difference between our $%
a_{2}^{\left( \mathrm{PF}\right) }+$ $a_{1}^{\left( \mathrm{PF}\right) }$
and the corresponding sum from~\cite{multi} is a $s$-exact modulo $d$
quantity. The associated component of antighost number zero, $a_{0}^{\left(
\mathrm{PF}\right) }$, is nevertheless the same in both formulations. As a
consequence, the object $a^{\left( \mathrm{PF}\right) }$ and the first-order
deformation in~\cite{multi} belong to the same cohomological class from $%
H^{0}\left( s|d\right) $.}. Consequently, it follows that $a^{\left( \mathrm{%
int}\right) }$ and $a^{\left( \mathrm{Dirac}\right) }$ are subject to some
separate equations
\begin{eqnarray}
sa^{\left( \mathrm{int}\right) } &=&\partial _{\mu }m^{\left( \mathrm{int}%
\right) \mu },  \label{PFDint} \\
sa^{\left( \mathrm{Dirac}\right) } &=&\partial _{\mu }m^{\left( \mathrm{Dirac%
}\right) \mu },  \label{PFDdir}
\end{eqnarray}
for some local $m^{\mu }$'s. In the sequel we analyze the general solutions
to these equations.

Since the Dirac field does not carry gauge symmetries of its own, it results
that the Dirac sector can only occur in antighost number one and zero, so,
without loss of generality, we take
\begin{equation}
a^{\left( \mathrm{int}\right) }=a_{0}^{\left( \mathrm{int}\right)
}+a_{1}^{\left( \mathrm{int}\right) }  \label{PFD3.21}
\end{equation}
in (\ref{PFDint}), where the components involved in the right-hand side of (%
\ref{PFD3.21}) are subject to the equations
\begin{eqnarray}
\gamma a_{1}^{\left( \mathrm{int}\right) } &=&0,  \label{PFD3.14a} \\
\delta a_{1}^{\left( \mathrm{int}\right) }+\gamma a_{0}^{\left( \mathrm{int}%
\right) } &=&\partial _{\mu }\stackrel{(0)}{m}^{\left(
\mathrm{int}\right) \mu }.  \label{PFD3.14b}
\end{eqnarray}
According to (\ref{PFD3.10}) in pure ghost number one and because $\omega
^{1}$ is spanned by
\[
\omega ^{1\Delta }=\left( \eta _{\mu },\;\partial _{[\mu }\eta _{\nu
]}\right) ,
\]
we infer that the most general expression of $a_{1}^{\left( \mathrm{int}%
\right) }$ as solution to the equation (\ref{PFD3.14a}), which complies with
all the general requirements imposed on the interacting theory (including
the preservation of the number of derivatives on each field with respect to
the free theory), is
\begin{eqnarray}
a_{1}^{\left( \mathrm{int}\right) } &=&\left( k_{1}\psi ^{*}\left( \partial
^{\alpha }\psi \right) +k_{1}^{\dagger }\left( \partial ^{\alpha }\bar{\psi}%
\right) \bar{\psi}^{*}+k_{2}\psi ^{*}\gamma ^{\alpha }\psi \right.  \nonumber
\\
&&\left. +k_{2}^{\dagger }\bar{\psi}\gamma ^{\alpha }\bar{\psi}%
^{*}+k_{3}\psi ^{*}\gamma ^{\alpha }\gamma ^{\mu }\left( \partial _{\mu
}\psi \right) +k_{3}^{\dagger }\left( \partial _{\mu }\bar{\psi}\right)
\gamma ^{\mu }\gamma ^{\alpha }\bar{\psi}^{*}\right) \eta _{\alpha }
\nonumber \\
&&+\left( k_{4}\bar{\psi}\left[ \gamma ^{\alpha },\gamma ^{\beta }\right]
\bar{\psi}^{*}-k_{4}^{\dagger }\psi ^{*}\left[ \gamma ^{\alpha },\gamma
^{\beta }\right] \psi \right) \partial _{[\alpha }\eta _{\beta ]}.
\label{PFD3.22}
\end{eqnarray}
Here, $\left( k_{j}\right) _{j=\overline{1,4}}$ are arbitrary complex
functions of $\bar{\psi}$ and $\psi $. If we represent them like
\begin{equation}
k_{j}=u_{j}+\mathrm{i}v_{j},\;j=\overline{1,4},  \label{PFD3.22a}
\end{equation}
with $u_{j}$ and $v_{j}$ real functions, then direct calculations, based on
the definitions (\ref{PFD12})--(\ref{PFD16}), lead to the elimination of
some of these functions from (\ref{PFD3.22}). For instance, the pieces
proportional with the real part of $k_{3}$ are
\begin{eqnarray}
&&u_{3}\left( \psi ^{*}\gamma ^{\alpha }\gamma ^{\mu }\left( \partial _{\mu
}\psi \right) +\left( \partial _{\mu }\bar{\psi}\right) \gamma ^{\mu }\gamma
^{\alpha }\bar{\psi}^{*}\right) \eta _{\alpha }  \nonumber \\
&=&s\left( -\mathrm{i}u_{3}\psi ^{*}\gamma ^{\alpha }\bar{\psi}^{*}\eta
_{\alpha }\right) -m\left( \left( \mathrm{i}u_{3}\right) \psi ^{*}\gamma
^{\alpha }\psi +\left( \mathrm{i}u_{3}\right) ^{\dagger }\bar{\psi}\gamma
^{\alpha }\bar{\psi}^{*}\right) \eta _{\alpha }.  \label{PFD3.23a}
\end{eqnarray}
However, we already have in $a_{1}^{\left( \mathrm{int}\right) }$ terms
proportional with $\psi ^{*}\gamma ^{\alpha }\psi \eta _{\alpha }$ and $\bar{%
\psi}\gamma ^{\alpha }\bar{\psi}^{*}\eta _{\alpha }$. So, if we set $%
k_{2}\rightarrow k_{2}^{\prime }=k_{2}-\mathrm{i}mu_{3}$ in (\ref{PFD3.22}),
then we can absorb the components proportional with $u_{3}$ into those
containing $k_{2}^{\prime }$ and $\left( k_{2}^{\prime }\right) ^{\dagger }$
since one can always remove the $s$-exact terms from $a^{\left( \mathrm{int}%
\right) }$ appearing in (\ref{PFD3.23a}) through a redefinition of the type (%
\ref{PFD3.1a}) corresponding to $n^{\mu }=0$. The above analysis leads to
the fact that we can safely take
\begin{equation}
u_{3}=0  \label{PFD3.24}
\end{equation}
in (\ref{PFD3.22}), without loss of independent contributions to $%
a_{1}^{\left( \mathrm{int}\right) }$. Strictly speaking, one may add to $%
a_{1}^{\left( \mathrm{int}\right) }$ given by (\ref{PFD3.22}) a term of the
type $\tilde{a}_{1}^{\left( \mathrm{int}\right) }=h^{*\mu \nu }\eta _{\mu
}F_{\nu }\left( \bar{\psi},\psi \right) $. On the one hand, we observe that
by applying $\delta $ on (\ref{PFD3.22}), then $a_{1}^{\left( \mathrm{int}%
\right) }$, if consistent, would lead to some $a_{0}^{\left( \mathrm{int}%
\right) }$ which contains a single field $h_{\mu \nu }$ (or one of its
first-order derivatives). On the other hand, from the expression of $\delta
\tilde{a}_{1}^{\left( \mathrm{int}\right) }$ we notice that if consistent,
it would give an $\tilde{a}_{0}^{\left( \mathrm{int}\right) }$ with two $%
h_{\mu \nu }$ (or one $h_{\mu \nu }$ and one of its first-order
derivatives). As a consequence, $\tilde{a}_{1}^{\left( \mathrm{int}\right) }$
must satisfy, independently of $a_{1}^{\left( \mathrm{int}\right) }$, an
equation of the type $\delta \tilde{a}_{1}^{\left( \mathrm{int}\right)
}+\gamma \tilde{a}_{0}^{\left( \mathrm{int}\right) }=\partial _{\mu }\rho
^{\mu }$. However, $\tilde{a}_{1}^{\left( \mathrm{int}\right) }$ produces a
consistent $\tilde{a}_{0}^{\left( \mathrm{int}\right) }$ if and only if $%
F_{\nu }\left( \bar{\psi},\psi \right) =\partial _{\nu }F\left( \bar{\psi}%
,\psi \right) $. The proof of the last statement can be found in Appendix %
\ref{appa}. Under these conditions, it is easy to see that
\begin{eqnarray}
\tilde{a}_{1}^{\left( \mathrm{int}\right) } &=&\partial _{\nu }\left(
h^{*\mu \nu }\eta _{\mu }F\left( \bar{\psi},\psi \right) \right) -\gamma
\left( \frac{1}{2}h^{*\mu \nu }h_{\mu \nu }F\left( \bar{\psi},\psi \right)
\right)  \nonumber \\
&&+s\left( \frac{1}{2}\eta ^{*\mu }\eta _{\mu }F\left( \bar{\psi},\psi
\right) \right) ,  \label{abc}
\end{eqnarray}
so $\tilde{a}_{1}^{\left( \mathrm{int}\right) }$ is trivial.

In order to analyze the solution $a_{0}^{\left( \mathrm{int}\right) }$ to
the equation (\ref{PFD3.14b}), it is useful to decompose $\delta
a_{1}^{\left( \mathrm{int}\right) }$ along the number of derivatives
\begin{equation}
\delta a_{1}^{\left( \mathrm{int}\right) }=\sum\limits_{k=0}^{2}\left(
\delta a_{1}^{\left( \mathrm{int}\right) }\right) _{k},  \label{PFD3.25}
\end{equation}
where $\left( \delta a_{1}^{\left( \mathrm{int}\right) }\right) _{k}$
denotes the piece with $k$-derivatives from $\delta a_{1}^{\left( \mathrm{int%
}\right) }$. According to this decomposition, it follows that each $\left(
\delta a_{1}^{\left( \mathrm{int}\right) }\right) _{k}$ should be written in
a $\gamma $-exact modulo $d$ form, such that (\ref{PFD3.14b}) is indeed
satisfied. Using the definitions (\ref{PFD12})--(\ref{PFD14}), we
consequently obtain
\begin{equation}
\left( \delta a_{1}^{\left( \mathrm{int}\right) }\right) _{0}=-m\left(
k_{2}+k_{2}^{\dagger }\right) \bar{\psi}\gamma ^{\alpha }\psi \eta _{\alpha
}.  \label{PFD3.26a}
\end{equation}
Due to the fact that the right-hand side of (\ref{PFD3.26a}) contains no
derivatives, it results that these terms neither reduce to a total
divergence nor can not produce $\gamma $-exact terms, so they must be made
to vanish
\begin{equation}
k_{2}+k_{2}^{\dagger }=0,  \label{PFD3.26b}
\end{equation}
which is equivalent, by means of (\ref{PFD3.22a}), with
\begin{equation}
u_{2}=0.  \label{PFD3.26c}
\end{equation}
The definitions (\ref{PFD12})--(\ref{PFD14}) and the result (\ref{PFD3.24})
together with (\ref{PFD3.26c}) further lead to
\begin{eqnarray}
&&\left( \delta a_{1}^{\left( \mathrm{int}\right) }\right) _{1}=-m\left(
\left( k_{1}-\frac{v_{2}}{m}+\mathrm{i}v_{3}\right) \bar{\psi}\left(
\partial ^{\alpha }\psi \right) +\left( k_{1}-\frac{v_{2}}{m}+\mathrm{i}%
v_{3}\right) ^{\dagger }\left( \partial ^{\alpha }\bar{\psi}\right) \psi
\right) \eta _{\alpha }  \nonumber \\
&&+\frac{m}{2}\left( \left( \frac{v_{2}}{m}+\mathrm{i}v_{3}\right) \left(
\partial _{\alpha }\bar{\psi}\right) \left[ \gamma ^{\alpha },\gamma ^{\beta
}\right] \psi -\left( \frac{v_{2}}{m}+\mathrm{i}v_{3}\right) ^{\dagger }\bar{%
\psi}\left[ \gamma ^{\alpha },\gamma ^{\beta }\right] \left( \partial
_{\alpha }\psi \right) \right) \eta _{\beta }  \nonumber \\
&&-m\left( k_{4}-k_{4}^{\dagger }\right) \bar{\psi}\left[ \gamma ^{\alpha
},\gamma ^{\beta }\right] \psi \partial _{\left[ \alpha \right. }\eta
_{\left. \beta \right] }.  \label{PFD3.27}
\end{eqnarray}
If we make the notations
\begin{equation}
U\left( \bar{\psi},\psi \right) =mk_{1}-v_{2}+\mathrm{i}mv_{3},  \label{def}
\end{equation}
\begin{equation}
V\left( \bar{\psi},\psi \right) =v_{2}+\mathrm{i}mv_{3},  \label{def1}
\end{equation}
then the formula (\ref{PFD3.27}) becomes
\begin{eqnarray}
&&\left( \delta a_{1}^{\left( \mathrm{int}\right) }\right) _{1}=\partial
^{\alpha }\left( -U^{\dagger }\bar{\psi}\psi \eta _{\alpha }+\frac{1}{2}V%
\bar{\psi}\left[ \gamma _{\alpha },\gamma _{\beta }\right] \psi \eta ^{\beta
}\right) +\gamma \left( \frac{1}{2}U^{\dagger }\bar{\psi}\psi h\right)
\nonumber \\
&&+\left( \left( U^{\dagger }-U\right) \bar{\psi}\left( \partial ^{\alpha
}\psi \right) +\frac{1}{2}\left( V+V^{\dagger }\right) \bar{\psi}\left[
\gamma ^{\alpha },\gamma ^{\beta }\right] \left( \partial _{\beta }\psi
\right) +\frac{1}{2}\bar{\psi}\left[ \gamma ^{\alpha },\gamma ^{\beta }%
\right] \psi \left( \partial _{\beta }V\right) \right.  \nonumber \\
&&\left. +\bar{\psi}\psi \left( \partial ^{\alpha }U^{\dagger }\right)
\right) \eta _{\alpha }-\left( \frac{1}{4}V+m\left( k_{4}-k_{4}^{\dagger
}\right) \right) \bar{\psi}\left[ \gamma ^{\alpha },\gamma ^{\beta }\right]
\psi \partial _{[\alpha }\eta _{\beta ]}.  \label{3.27}
\end{eqnarray}
The right-hand side from (\ref{3.27}) is $\gamma $-exact modulo $d$ if the
functions $U$ and $V$\ satisfy the equations
\begin{eqnarray}
&&\left( U^{\dagger }-U\right) \bar{\psi}\left( \partial ^{\alpha }\psi
\right) +\bar{\psi}\psi \left( \partial ^{\alpha }U^{\dagger }\right) +\frac{%
1}{2}\bar{\psi}\left[ \gamma ^{\alpha },\gamma ^{\beta }\right] \psi \left(
\partial _{\beta }V\right)  \nonumber \\
&&+\frac{1}{2}\left( V+V^{\dagger }\right) \bar{\psi}\left[ \gamma ^{\alpha
},\gamma ^{\beta }\right] \left( \partial _{\beta }\psi \right) =\partial
_{\beta }P^{\beta \alpha },  \label{x1}
\end{eqnarray}
where
\begin{equation}
\frac{1}{2}\left( P^{\alpha \beta }-P^{\beta \alpha }\right) =-\left( \frac{1%
}{2}V+2m\left( k_{4}-k_{4}^{\dagger }\right) \right) \bar{\psi}\left[ \gamma
^{\alpha },\gamma ^{\beta }\right] \psi .  \label{x2}
\end{equation}
By direct computation we find that the left-hand side of (\ref{x1}) reduces
to a total derivative if
\begin{equation}
-U\bar{\psi}\left( \partial ^{\alpha }\psi \right) -U^{\dagger }\left(
\partial ^{\alpha }\bar{\psi}\right) \psi =\frac{1}{2}\partial _{\beta
}\left( P^{\beta \alpha }+P^{\alpha \beta }-2\sigma ^{\alpha \beta
}U^{\dagger }\bar{\psi}\psi \right) ,  \label{x3}
\end{equation}
\begin{eqnarray}
&&-\frac{1}{2}V\left( \partial _{\beta }\bar{\psi}\right) \left[ \gamma
^{\alpha },\gamma ^{\beta }\right] \psi +\frac{1}{2}V^{\dagger }\bar{\psi}%
\left[ \gamma ^{\alpha },\gamma ^{\beta }\right] \left( \partial _{\beta
}\psi \right)  \nonumber \\
&=&\frac{1}{2}\partial _{\beta }\left( P^{\beta \alpha }-P^{\alpha \beta }-V%
\bar{\psi}\left[ \gamma ^{\alpha },\gamma ^{\beta }\right] \psi \right) .
\label{x4}
\end{eqnarray}
Now, the left-hand side from (\ref{x3}) is a total derivative if
\begin{equation}
U=U^{\dagger }  \label{x5}
\end{equation}
and, in addition, $U$ is a polynomial in $\bar{\psi}\psi $ with real
coefficients. In this situation we have that
\begin{equation}
-U\bar{\psi}\left( \partial ^{\alpha }\psi \right) -U^{\dagger }\left(
\partial ^{\alpha }\bar{\psi}\right) \psi =\partial ^{\alpha }W,
\label{yzw1}
\end{equation}
where the function $W$ is defined via the relation
\begin{equation}
U=-\frac{dW}{d\left( \bar{\psi}\psi \right) },  \label{y1}
\end{equation}
such that (\ref{x3}) can be written like
\begin{equation}
\partial _{\beta }\left( P^{\beta \alpha }+P^{\alpha \beta }-2\sigma
^{\alpha \beta }\left( U\bar{\psi}\psi +W\right) \right) =0.  \label{y2}
\end{equation}
Since the quantity $P^{\beta \alpha }+P^{\alpha \beta }-2\sigma ^{\alpha
\beta }\left( U\bar{\psi}\psi +W\right) $ contains no derivatives, from (\ref%
{y2}) we obtain that
\begin{equation}
P^{\beta \alpha }+P^{\alpha \beta }=2\sigma ^{\alpha \beta }\left( U\bar{\psi%
}\psi +W\right) .  \label{x6}
\end{equation}
Inserting (\ref{x2}) in (\ref{x4}), we arrive at
\begin{eqnarray}
&&-\frac{1}{2}V\left( \partial _{\beta }\bar{\psi}\right) \left[ \gamma
^{\alpha },\gamma ^{\beta }\right] \psi +\frac{1}{2}V^{\dagger }\bar{\psi}%
\left[ \gamma ^{\alpha },\gamma ^{\beta }\right] \left( \partial _{\beta
}\psi \right)  \nonumber \\
&=&2m\partial _{\beta }\left( \left( k_{4}-k_{4}^{\dagger }\right) \bar{\psi}%
\left[ \gamma ^{\alpha },\gamma ^{\beta }\right] \psi \right) .  \label{x7}
\end{eqnarray}
At this stage we observe that the left-hand side of the previous formula
leads to a total derivative if $V$ is a purely imaginary constant
\begin{equation}
V=\mathrm{const},\;V+V^{\dagger }=0,  \label{y3}
\end{equation}
in which case the relation (\ref{x7}) takes the form
\begin{equation}
\partial _{\beta }\left( -\frac{1}{2}V\bar{\psi}\left[ \gamma ^{\alpha
},\gamma ^{\beta }\right] \psi \right) =2m\partial _{\beta }\left( \left(
k_{4}-k_{4}^{\dagger }\right) \bar{\psi}\left[ \gamma ^{\alpha },\gamma
^{\beta }\right] \psi \right) ,  \label{y4}
\end{equation}
and thus
\begin{equation}
V=-4m\left( k_{4}-k_{4}^{\dagger }\right) .  \label{x8}
\end{equation}
Relying on the last result, by means of (\ref{x2}) we obtain
\begin{equation}
P^{\alpha \beta }-P^{\beta \alpha }=0,  \label{x9}
\end{equation}
such that (\ref{x6}) gives
\begin{equation}
P^{\alpha \beta }=\sigma ^{\alpha \beta }\left( U\bar{\psi}\psi +W\right) .
\label{y5}
\end{equation}
Let us analyze the results deduced so far. The relations (\ref{def1}) and (%
\ref{y3}) allow us to state that $v_{3}$ must be a true constant
\begin{equation}
v_{3}=\mathrm{const}_{1},  \label{yw1}
\end{equation}
while $v_{2}$ must vanish
\begin{equation}
v_{2}=0.  \label{x10}
\end{equation}
In the meantime, the formula (\ref{x8}) implies that
\begin{equation}
v_{4}=-\frac{v_{3}}{8}.  \label{x11}
\end{equation}
Using (\ref{def}), (\ref{x5}) and (\ref{x10}) we arrive at
\begin{equation}
v_{1}=-v_{3},  \label{x12}
\end{equation}
such that
\begin{equation}
U=mu_{1}.  \label{y7}
\end{equation}
Introducing all the above results into (\ref{3.27}), we infer that
\begin{eqnarray}
\left( \delta a_{1}^{\left( \mathrm{int}\right) }\right) _{1} &=&\partial
_{\alpha }\left( W\left( \bar{\psi}\psi \right) \eta ^{\alpha }+\frac{%
\mathrm{i}mv_{3}}{2}\bar{\psi}\left[ \gamma ^{\alpha },\gamma ^{\beta }%
\right] \psi \eta _{\beta }\right)  \nonumber \\
&&-\gamma \left( \frac{1}{2}W\left( \bar{\psi}\psi \right) h\right) .
\label{x13}
\end{eqnarray}
Taking into account the results (\ref{x10})--(\ref{x12}), the pieces
containing two derivatives from $\delta a_{1}^{\left( \mathrm{int}\right) }$
can be written like
\begin{eqnarray}
\left( \delta a_{1}^{\left( \mathrm{int}\right) }\right) _{2} &=&\partial
_{\mu }\left( \mathrm{i}u_{1}\bar{\psi}\left( \gamma ^{\alpha }\left(
\partial _{\alpha }\psi \right) \eta ^{\mu }-\gamma ^{\mu }\left( \partial
^{\alpha }\psi \right) \eta _{\alpha }\right) \right.  \nonumber \\
&&+\left( \mathrm{i}u_{4}-\frac{v_{3}}{8}\right) \bar{\psi}\gamma ^{\mu }%
\left[ \gamma ^{\alpha },\gamma ^{\beta }\right] \psi \partial _{[\alpha
}\eta _{\beta ]}  \nonumber \\
&&\left. +\frac{v_{3}}{2}\bar{\psi}\left( \gamma ^{\alpha }\left[ \gamma
^{\mu },\gamma ^{\beta }\right] -\gamma ^{\mu }\left[ \gamma ^{\alpha
},\gamma ^{\beta }\right] \right) \left( \partial _{\alpha }\psi \right)
\eta _{\beta }\right)  \nonumber \\
&&+\gamma \left( -\frac{\mathrm{i}u_{1}}{2}\bar{\psi}\left( \gamma ^{\mu
}\left( \partial _{\mu }\psi \right) h-\gamma ^{\alpha }\left( \partial
^{\beta }\psi \right) h_{\alpha \beta }\right) \right.  \nonumber \\
&&-\left( \mathrm{i}u_{4}-\frac{v_{3}}{8}\right) \bar{\psi}\gamma ^{\mu }%
\left[ \gamma ^{\alpha },\gamma ^{\beta }\right] \psi \partial _{[\alpha
}h_{\beta ]\mu }  \nonumber \\
&&\left. +\frac{v_{3}}{4}\bar{\psi}\gamma ^{\mu }\left[ \gamma ^{\alpha
},\gamma ^{\beta }\right] \left( \partial _{\alpha }\psi \right) h_{\beta
\mu }\right)  \nonumber \\
&&+\left( \mathrm{i}\left( \partial _{\mu }u_{1}\right) \bar{\psi}\gamma
^{\mu }\left( \partial ^{\alpha }\psi \right) -\mathrm{i}\left( \partial
^{\alpha }u_{1}\right) \bar{\psi}\gamma ^{\mu }\left( \partial _{\mu }\psi
\right) \right.  \nonumber \\
&&\left. +2v_{3}\left( \left( \partial _{\mu }\bar{\psi}\right) \gamma ^{\mu
}\left( \partial ^{\alpha }\psi \right) -\left( \partial _{\mu }\bar{\psi}%
\right) \gamma ^{\alpha }\left( \partial ^{\mu }\psi \right) \right) \right)
\eta _{\alpha }  \nonumber \\
&&+\left( \frac{\mathrm{i}}{2}\left( u_{1}+16u_{4}+\mathrm{i}v_{3}\right)
\bar{\psi}\gamma ^{\alpha }\partial ^{\beta }\psi \right.  \nonumber \\
&&\left. -\bar{\psi}\gamma ^{\mu }\left[ \gamma ^{\alpha },\gamma ^{\beta }%
\right] \left( \psi \mathrm{i}\left( \partial _{\mu }u_{4}\right) +\frac{%
v_{3}}{4}\left( \partial _{\mu }\psi \right) \right) \right) \partial
_{[\alpha }\eta _{\beta ]}.  \label{x14}
\end{eqnarray}
In consequence, $\left( \delta a_{1}^{\left( \mathrm{int}\right) }\right)
_{2}$ is $\gamma $-exact modulo $d$ if
\begin{eqnarray}
v_{3} &=&0,  \label{x15a} \\
u_{1}\left( \bar{\psi}\psi \right) +16u_{4}\left( \bar{\psi},\psi \right) +%
\mathrm{i}v_{3} &=&0,  \label{x15b} \\
\partial _{\mu }u_{4}\left( \bar{\psi},\psi \right) &=&\partial _{\mu
}u_{1}\left( \bar{\psi}\psi \right) =0.  \label{x15c}
\end{eqnarray}
By means of (\ref{x15b})--(\ref{x15c}) we get that the functions $u_{1}$ and
$u_{4}$ are some constants
\begin{equation}
u_{1}=\mathrm{const}_{2},\;u_{4}=\mathrm{const}_{3},  \label{y10}
\end{equation}
related via the formula
\begin{equation}
u_{4}=-\frac{u_{1}}{16}.  \label{x16}
\end{equation}
As $u_{1}$ is constant, from (\ref{y1}) and (\ref{y7}) we find that
\begin{equation}
W=-u_{1}m\bar{\psi}\psi ,  \label{y8}
\end{equation}
such that (\ref{x13}) becomes
\begin{equation}
\left( \delta a_{1}^{\left( \mathrm{int}\right) }\right) _{1}=\partial
_{\alpha }\left( -u_{1}m\bar{\psi}\psi \eta ^{\alpha }\right) +\gamma \left(
\frac{u_{1}}{2}m\bar{\psi}\psi h\right) .  \label{y9}
\end{equation}
Introducing the results (\ref{x15a}) and (\ref{x16}) in (\ref{x14}), it
follows that
\begin{eqnarray}
\left( \delta a_{1}^{\left( \mathrm{int}\right) }\right) _{2} &=&\partial
_{\mu }\left( \mathrm{i}u_{1}\bar{\psi}\left( \gamma ^{\alpha }\left(
\partial _{\alpha }\psi \right) \eta ^{\mu }-\gamma ^{\mu }\left( \partial
^{\alpha }\psi \right) \eta _{\alpha }\right. \right.  \nonumber \\
&&\left. \left. -\frac{1}{16}\gamma ^{\mu }\left[ \gamma ^{\alpha },\gamma
^{\beta }\right] \psi \partial _{[\alpha }\eta _{\beta ]}\right) \right)
\nonumber \\
&&+\gamma \left( -\frac{\mathrm{i}u_{1}}{2}\bar{\psi}\left( \gamma ^{\mu
}\left( \partial _{\mu }\psi \right) h-\gamma ^{\alpha }\left( \partial
^{\beta }\psi \right) h_{\alpha \beta }\right. \right.  \nonumber \\
&&\left. \left. -\frac{1}{8}\gamma ^{\mu }\left[ \gamma ^{\alpha },\gamma
^{\beta }\right] \psi \partial _{[\alpha }h_{\beta ]\mu }\right) \right) .
\label{x17}
\end{eqnarray}

Putting together the formulas (\ref{PFD3.22a}), (\ref{PFD3.24}), (\ref%
{PFD3.26c}), (\ref{x10})--(\ref{x12}), (\ref{x15a}), and (\ref{y10})--(\ref%
{x16}), we conclude that the most general expression of $a_{1}^{\left(
\mathrm{int}\right) }$ that produces a consistent component in antighost
number zero as solution to the equation (\ref{PFD3.14b}) and complies with
all the general requirements imposed at the beginning of this section can be
expressed in terms of a single arbitrary, real constant, $u_{1}$. From now
on we will denote this constant by $k$, such that the resulting $%
a_{1}^{\left( \mathrm{int}\right) }$ becomes
\begin{eqnarray}
a_{1}^{\left( \mathrm{int}\right) } &=&-k\left( \frac{1}{16}\left( \bar{\psi}%
\left[ \gamma ^{\alpha },\gamma ^{\beta }\right] \bar{\psi}^{*}-\psi ^{*}%
\left[ \gamma ^{\alpha },\gamma ^{\beta }\right] \psi \right) \partial
_{[\alpha }\eta _{\beta ]}\right.  \nonumber \\
&&\left. -\left( \psi ^{*}\left( \partial ^{\alpha }\psi \right) +\left(
\partial ^{\alpha }\bar{\psi}\right) \bar{\psi}^{*}\right) \eta _{\alpha
}\right) .  \label{PFD3.35}
\end{eqnarray}
Then, using (\ref{PFD3.14b}), (\ref{PFD3.25}), (\ref{PFD3.26a})--(\ref%
{PFD3.26b}), (\ref{y9}) and (\ref{x17}) we find that the corresponding $%
a_{0}^{\left( \mathrm{int}\right) }$ is
\begin{eqnarray}
a_{0}^{\left( \mathrm{int}\right) } &=&\frac{k}{2}\left( \bar{\psi}\left(
\mathrm{i}\gamma ^{\mu }\left( \partial _{\mu }\psi \right) -m\psi \right) h-%
\mathrm{i}\bar{\psi}\gamma ^{\alpha }\left( \partial ^{\beta }\psi \right)
h_{\alpha \beta }\right.  \nonumber \\
&&\left. -\frac{\mathrm{i}}{8}\bar{\psi}\gamma ^{\mu }\left[ \gamma ^{\alpha
},\gamma ^{\beta }\right] \psi \partial _{[\alpha }h_{\beta ]\mu }\right) +%
\bar{a}_{0}^{\left( \mathrm{int}\right) }.  \label{PFD3.36}
\end{eqnarray}

In the above, $\bar{a}_{0}^{\left( \mathrm{int}\right) }$ represents the
general local solution to the homogeneous equation
\begin{equation}
\gamma \bar{a}_{0}^{\left( \mathrm{int}\right) }=\partial _{\mu }\bar{m}%
^{\left( \mathrm{int}\right) \mu },  \label{PFD3.39}
\end{equation}
for some local $\bar{m}^{\left( \mathrm{int}\right) \mu }$. Such solutions
correspond to $\bar{a}_{1}^{\left( \mathrm{int}\right) }=0$ and thus they
cannot deform either the gauge algebra or the gauge transformations, but
simply the lagrangian at order one in the coupling constant. There are two
main types of solutions to (\ref{PFD3.39}). The first one corresponds to $%
\bar{m}^{\left( \mathrm{int}\right) \mu }=0$ and is given by
gauge-invariant, non-integrated densities constructed from the original
fields and their spacetime derivatives. According to (\ref{PFD3.10}) for
both pure ghost and antighost numbers equal to zero, they are given by $\bar{%
a}_{0}^{\prime \left( \mathrm{int}\right) }=\bar{a}_{0}^{\prime \left(
\mathrm{int}\right) }\left( \left[ \psi \right] ,\left[ \bar{\psi}\right] ,%
\left[ K_{\mu \nu \alpha \beta }\right] \right) $, up to the conditions that
they effectively describe cross-couplings between the two types of fields
and cannot be written in a divergence-like form. Unfortunately, this type of
solutions must depend on the linearized Riemann tensor (and possibly of its
derivatives) in order to provide cross-couplings, and thus would lead to
terms with at least two derivatives of the Dirac spinors. So, by virtue of
the derivative order assumption, they must be discarded by setting $\bar{a}%
_{0}^{\prime \left( \mathrm{int}\right) }=0$. The second kind of solutions
is associated with $\bar{m}^{\left( \mathrm{int}\right) \mu }\neq 0$ in (\ref%
{PFD3.39}), being understood that they lead to cross-interactions, cannot be
written in a divergence-like form and contain at most one derivative of the
fields. Consequently, we obtain that
\begin{eqnarray}
\gamma \bar{a}_{0}^{\left( \mathrm{int}\right) } &=&\partial _{\rho }\left(
\frac{\partial \bar{a}_{0}^{\left( \mathrm{int}\right) }}{\partial \left(
\partial _{\rho }h_{\mu \nu }\right) }\partial _{(\mu }\eta _{\nu )}\right) +%
\frac{\delta \bar{a}_{0}^{\left( \mathrm{int}\right) }}{\delta h_{\mu \nu }}%
\partial _{(\mu }\eta _{\nu )}  \nonumber \\
&=&\partial _{\rho }\left( \frac{\partial \bar{a}_{0}^{\left( \mathrm{int}%
\right) }}{\partial \left( \partial _{\rho }h_{\mu \nu }\right) }\partial
_{(\mu }\eta _{\nu )}+2\frac{\delta \bar{a}_{0}^{\left( \mathrm{int}\right) }%
}{\delta h_{\rho \mu }}\eta _{\mu }\right) -2\partial _{\rho }\left( \frac{%
\delta \bar{a}_{0}^{\left( \mathrm{int}\right) }}{\delta h_{\rho \mu }}%
\right) \eta _{\mu }.  \label{PFD3.42}
\end{eqnarray}
Thus, this $\bar{a}_{0}^{\left( \mathrm{int}\right) }$ fulfills (\ref%
{PFD3.39}) if and only if its Euler-Lagrange derivatives with respect to the
Pauli-Fierz fields satisfy the equations
\begin{equation}
\partial _{\rho }\left( \frac{\delta \bar{a}_{0}^{\left( \mathrm{int}\right)
}}{\delta h_{\rho \mu }}\right) =0.  \label{PFD3.42a}
\end{equation}
Since $\bar{a}_{0}^{\left( \mathrm{int}\right) }$ may contain at most one
derivative, it follows that the solution to (\ref{PFD3.42a}) reads as
\begin{equation}
\frac{\delta \bar{a}_{0}^{\left( \mathrm{int}\right) }}{\delta h_{\mu \nu }}%
=\partial _{\rho }D^{\rho \mu \nu },  \label{PFD3.43}
\end{equation}
where $D^{\rho \mu \nu }$ contains no derivatives and is antisymmetric in
its first two indices\footnote{%
Strictly speaking, we might have added to the right-hand side of (\ref%
{PFD3.43}) the contribution $c\sigma ^{\mu \nu }$, with $c$ an arbitrary
real constant. It would not have led to cross-interactions, but to the
cosmological term $ch$, which has already been considered in $a^{\left(
\mathrm{PF}\right) }$.}
\begin{equation}
D^{\rho \mu \nu }=-D^{\mu \rho \nu }.  \label{PFD3.44a}
\end{equation}
We insist on the fact that a solution of the type $\delta \bar{a}%
_{0}^{\left( \mathrm{int}\right) }/\delta h_{\mu \nu }=\partial _{\alpha
}\partial _{\beta }D^{\mu \alpha \nu \beta }$, with $D^{\mu \alpha \nu \beta
}$ possessing the symmetry properties of the Riemann tensor, is not allowed
in our case due to the hypothesis on the derivative order, and hence (\ref%
{PFD3.43}) is the most general solution to the equation (\ref{PFD3.42a}).
Moreover, from (\ref{PFD3.43}) we have that $D^{\rho \mu \nu }$ must be
symmetric in its last two indices
\begin{equation}
D^{\rho \mu \nu }=D^{\rho \nu \mu }.  \label{PFD3.44b}
\end{equation}
The properties (\ref{PFD3.44a}) and (\ref{PFD3.44b}) imply that
\begin{eqnarray}
D^{\rho \mu \nu } &=&-D^{\mu \rho \nu }=-D^{\mu \nu \rho }=D^{\nu \mu \rho }
\nonumber \\
&=&D^{\nu \rho \mu }=-D^{\rho \nu \mu }=-D^{\rho \mu \nu },  \label{PFD3.45}
\end{eqnarray}
and hence $D^{\rho \mu \nu }=0$. Consequently, (\ref{PFD3.43}) yields
\begin{equation}
\frac{\delta \bar{a}_{0}^{\left( \mathrm{int}\right) }}{\delta h_{\mu \nu }}%
=0,  \label{PFD3.46}
\end{equation}
and thus we can write that $\bar{a}_{0}^{\left( \mathrm{int}\right)
}=L\left( \left[ \psi \right] ,\left[ \bar{\psi}\right] \right) +\partial
_{\mu }g^{\mu }\left( \psi ,\bar{\psi},h_{\alpha \beta }\right) $. Since we
are interested only in cross-interactions, we must set $L=0$. At this stage
we remain with the trivial solutions
\begin{equation}
\bar{a}_{0}^{\left( \mathrm{int}\right) }=\partial _{\mu }g^{\mu }\left(
\psi ,\bar{\psi},h_{\alpha \beta }\right) ,  \label{PFD3.47}
\end{equation}
which can be completely removed from the first-order deformation via a
transformation of the form (\ref{PFD3.1a}) with $b=0$. In conclusion, we can
take, without loss of generality
\begin{equation}
\bar{a}_{0}^{\left( \mathrm{int}\right) }=0  \label{PFD3.48}
\end{equation}
in the solution (\ref{PFD3.36}). As a consequence of the above discussion,
we can state that the antighost number zero component of $a^{\left( \mathrm{%
int}\right) }$ reads as
\begin{eqnarray}
a_{0}^{\left( \mathrm{int}\right) } &=&\frac{k}{2}\left( \bar{\psi}\left(
\mathrm{i}\gamma ^{\mu }\left( \partial _{\mu }\psi \right) -m\psi \right) h-%
\mathrm{i}\bar{\psi}\gamma ^{\alpha }\left( \partial ^{\beta }\psi \right)
h_{\alpha \beta }\right.  \nonumber \\
&&\left. -\frac{\mathrm{i}}{8}\bar{\psi}\gamma ^{\mu }\left[ \gamma ^{\alpha
},\gamma ^{\beta }\right] \psi \partial _{[\alpha }h_{\beta ]\mu }\right) .
\label{AB1}
\end{eqnarray}
After some computation, we find that
\begin{eqnarray}
a_{1}^{\left( \mathrm{int}\right) }+a_{0}^{\left( \mathrm{int}\right) }
&=&-k\left( \frac{1}{16}\left( \bar{\psi}\left[ \gamma ^{\alpha },\gamma
^{\beta }\right] \bar{\psi}^{*}-\psi ^{*}\left[ \gamma ^{\alpha },\gamma
^{\beta }\right] \psi \right) \partial _{[\alpha }\eta _{\beta ]}\right.
\nonumber \\
&&\left. -\frac{1}{2}\left( \psi ^{*}\left( \partial ^{\alpha }\psi \right)
+\left( \partial ^{\alpha }\bar{\psi}\right) \bar{\psi}^{*}-\left( \partial
^{\alpha }\psi ^{*}\right) \psi -\bar{\psi}\left( \partial ^{\alpha }\bar{%
\psi}^{*}\right) \right) \eta _{\alpha }\right)  \nonumber \\
&&-\frac{k}{2}\Theta ^{\mu \nu }h_{\mu \nu }+s\Lambda +\partial _{\mu
}v^{\mu },  \label{ABN}
\end{eqnarray}
where
\begin{equation}
\Theta ^{\mu \nu }=\frac{\mathrm{i}}{2}\left( \bar{\psi}\gamma ^{\left( \mu
\right. }\partial ^{\left. \nu \right) }\psi -\left( \partial ^{\left( \mu
\right. }\bar{\psi}\right) \gamma ^{\left. \nu \right) }\psi \right) ,
\label{AB2}
\end{equation}
represents the stress-energy tensor of the Dirac field, while $\Lambda $ is
given by
\begin{equation}
\Lambda =-\frac{k}{4}\left( \psi ^{*}\psi +\bar{\psi}\bar{\psi}^{*}\right) h.
\label{AB3}
\end{equation}
Obviously, the term $s\Lambda +\partial _{\mu }v^{\mu }$ from (\ref{ABN}) is
cohomologically trivial, and hence can be discarded. Thus, the coupling
between a Dirac field and one graviton at the first order in the deformation
parameter takes the form $\Theta ^{\mu \nu }h_{\mu \nu }$. We cannot stress
enough that is not an assumption, but follows entirely from the deformation
approach developed here. However, for subsequent purposes it is useful to
work with the expressions (\ref{PFD3.35}) and (\ref{AB1}) of $a_{0}^{\left(
\mathrm{int}\right) }$ and $a_{1}^{\left( \mathrm{int}\right) }$.

Finally, we analyze the component $a^{\left( \mathrm{Dirac}\right) }$ from (%
\ref{PFD3.12a}). As the Dirac action from (\ref{fract}) has no non-trivial
gauge invariance, it follows that $a^{\left( \mathrm{Dirac}\right) }$ can
only reduce to its component of antighost number zero
\begin{equation}
a^{\left( \mathrm{Dirac}\right) }=a_{0}^{\left( \mathrm{Dirac}\right)
}\left( \left[ \psi \right] ,\left[ \bar{\psi}\right] \right) ,
\label{PFD3.49}
\end{equation}
which is automatically solution to the equation $sa^{\left( \mathrm{Dirac}%
\right) }\equiv \gamma a_{0}^{\left( \mathrm{Dirac}\right) }=0$. It comes
from $a_{1}^{\left( \mathrm{Dirac}\right) }=0$ and does not deform the gauge
transformations (\ref{PFD5}), but merely modifies the Dirac action. The
condition that $a_{0}^{\left( \mathrm{Dirac}\right) }$ is of maximum
derivative order equal to one is translated into
\begin{equation}
a_{0}^{\left( \mathrm{Dirac}\right) }=f\left( \bar{\psi},\psi \right)
+\left( \partial _{\mu }\bar{\psi}\right) g_{1}^{\mu }\left( \bar{\psi},\psi
\right) +g_{2}^{\mu }\left( \bar{\psi},\psi \right) \left( \partial _{\mu
}\psi \right) ,  \label{PFD3.50}
\end{equation}
where $f$, $g_{1}^{\mu }$ and $g_{2}^{\mu }$ are polynomials in the
undifferentiated spinor fields (since they anticommute). The first
polynomial is a scalar (bosonic and real), while the one-tensors $g_{1}^{\mu
}$ and $g_{2}^{\mu }$ are fermionic and spinor-like. They are related via
the relation
\begin{equation}
\left( g_{1}^{\mu }\right) ^{\dagger }\gamma _{0}=g_{2}^{\mu }
\label{PFD3.51}
\end{equation}
in order to ensure that $a_{0}^{\left( \mathrm{Dirac}\right) }$ is indeed a
scalar.

\subsection{Second-order deformation}

In the previous part of the paper we have seen that the first-order
deformation of the theory can be written like the sum between the
first-order deformation of the master equation for the Pauli-Fierz theory $%
S_{1}^{\left( \mathrm{PF}\right) }$, and the `interacting' part $%
S_{1}^{\left( \mathrm{int}\right) }$
\begin{equation}
S_{1}^{\left( \mathrm{int}\right) }=\int d^{4}x\left( a_{1}^{\left( \mathrm{%
int}\right) }+a_{0}^{\left( \mathrm{int}\right) }+a_{0}^{\left( \mathrm{Dirac%
}\right) }\right) ,  \label{PFD4.1}
\end{equation}
with $a_{1}^{\left( \mathrm{int}\right) }$, $a_{0}^{\left( \mathrm{int}%
\right) }$ and $a_{0}^{\left( \mathrm{Dirac}\right) }$ respectively given by
(\ref{PFD3.35}), (\ref{PFD3.36}) and (\ref{PFD3.50}).

As shown in Appendix \ref{appb}, the first-order deformation is consistent
at order two in the coupling constant if the constant $k$ that parametrizes
both $a_{1}^{\left( \mathrm{int}\right) }$ and $a_{0}^{\left( \mathrm{int}%
\right) }$ is equal to unit
\begin{equation}
k=1,  \label{w1}
\end{equation}
and the functions appearing in $a_{0}^{\left( \mathrm{Dirac}\right) }$ are
of the form
\begin{equation}
f\left( \bar{\psi},\psi \right) =M\left( \bar{\psi}\psi \right)
,\;g_{1}^{\mu }=0,\;g_{2}^{\mu }=0,  \label{w2}
\end{equation}
with $M\left( \bar{\psi}\psi \right) $ a polynomial in $\bar{\psi}\psi $.
Under these circumstances, we have that $S_{2}=S_{2}^{\left( \mathrm{PF}%
\right) }+S_{2}^{\left( \mathrm{int}\right) }$, where $S_{2}^{\left( \mathrm{%
PF}\right) }$ can be deduced from~\cite{multi}, and
\begin{eqnarray}
S_{2}^{\left( \mathrm{int}\right) } &=&\int d^{4}x\left( -\frac{1}{2}\left(
\psi ^{*}\left( \partial ^{\alpha }\psi \right) +\left( \partial ^{\alpha }%
\bar{\psi}\right) \bar{\psi}^{*}\right) \eta ^{\beta }h_{\alpha \beta }+%
\frac{1}{32}\left( \bar{\psi}\left[ \gamma ^{\alpha },\gamma ^{\beta }\right]
\bar{\psi}^{*}\right. \right.  \nonumber \\
&&\left. -\psi ^{*}\left[ \gamma ^{\alpha },\gamma ^{\beta }\right] \psi
\right) \left( \eta ^{\sigma }\partial _{[\alpha }h_{\beta ]\sigma }-\frac{1%
}{2}h_{\sigma [\alpha }\left( \partial _{\beta ]}\eta ^{\sigma }-\partial
^{\sigma }\eta _{\beta ]}\right) \right)  \nonumber \\
&&-\frac{\mathrm{i}}{4}\bar{\psi}\gamma ^{\mu }\left( \partial ^{\nu }\psi
\right) \left( hh_{\mu \nu }-\frac{3}{2}h_{\mu \sigma }h_{\nu }^{\sigma
}\right) -\frac{1}{4}\left( \bar{\psi}\mathrm{i}\gamma ^{\mu }\left(
\partial _{\mu }\psi \right) -m\bar{\psi}\psi \right) \times  \nonumber \\
&&\times \left( h_{\alpha \beta }h^{\alpha \beta }-\frac{1}{2}h^{2}\right) -%
\frac{\mathrm{i}}{32}\bar{\psi}\gamma ^{\mu }\left[ \gamma ^{\alpha },\gamma
^{\beta }\right] \psi \left( h\partial _{[\alpha }h_{\beta ]\mu }\right.
\nonumber \\
&&\left. \left. -h_{\mu }^{\sigma }\partial _{[\alpha }h_{\beta ]\sigma
}+h_{\alpha }^{\sigma }\left( 2\partial _{[\beta }h_{\sigma ]\mu }+\partial
_{\mu }h_{\beta \sigma }\right) \right) +\frac{1}{2}M\left( \bar{\psi}\psi
\right) h\right) .  \label{w3}
\end{eqnarray}
The concrete expression of $S_{2}^{\left( \mathrm{int}\right) }$ is inferred
also in Appendix \ref{appb}. Making use of (\ref{w1})--(\ref{w2}), it
results that $S_{1}^{\left( \mathrm{int}\right) }$ takes the final form
\begin{eqnarray}
S_{1}^{\left( \mathrm{int}\right) } &=&\int d^{4}x\left( -\frac{1}{16}\left(
\bar{\psi}\left[ \gamma ^{\alpha },\gamma ^{\beta }\right] \bar{\psi}%
^{*}-\psi ^{*}\left[ \gamma ^{\alpha },\gamma ^{\beta }\right] \psi \right)
\partial _{[\alpha }\eta _{\beta ]}\right.  \nonumber \\
&&+\left( \psi ^{*}\left( \partial ^{\alpha }\psi \right) +\left( \partial
^{\alpha }\bar{\psi}\right) \bar{\psi}^{*}\right) \eta _{\alpha }  \nonumber
\\
&&+\frac{1}{2}\left( \bar{\psi}\left( \mathrm{i}\gamma ^{\mu }\left(
\partial _{\mu }\psi \right) -m\psi \right) h-\mathrm{i}\bar{\psi}\gamma
^{\alpha }\left( \partial ^{\beta }\psi \right) h_{\alpha \beta }\right.
\nonumber \\
&&\left. \left. -\frac{\mathrm{i}}{8}\bar{\psi}\gamma ^{\mu }\left[ \gamma
^{\alpha },\gamma ^{\beta }\right] \psi \partial _{[\alpha }h_{\beta ]\mu
}\right) +M\left( \bar{\psi}\psi \right) \right) .  \label{w4}
\end{eqnarray}

\section{Vierbein versus Pauli-Fierz formulation of spin-two field theory}

In this section we correlate the linearized versions of first- and
second-order formulations of spin-two field theory via the local BRST
cohomology. In view of this, we start from the first-order formulation of
spin-two field theory
\begin{eqnarray}
S\left[ e_{a}^{\;\;\mu },\omega _{\mu ab}\right] &=&-\frac{1}{\lambda }\int
d^{4}x\left( \omega _{\nu }^{\;\;ab}\partial _{\mu }\left( ee_{a}^{\;\;\mu
}e_{b}^{\;\;\nu }\right) -\omega _{\mu }^{\;\;ab}\partial _{\nu }\left(
ee_{a}^{\;\;\mu }e_{b}^{\;\;\nu }\right) \right.  \nonumber \\
&&\left. +\frac{1}{2}ee_{a}^{\;\;\mu }e_{b}^{\;\;\nu }\left( \omega _{\mu
}^{\;\;ac}\omega _{\nu \;\;\;c}^{\;\;b}-\omega _{\nu }^{\;\;ac}\omega _{\mu
\;\;\;c}^{\;\;b}\right) \right) ,  \label{xx1}
\end{eqnarray}
where $e_{a}^{\;\;\mu }$ is the vierbein field and $\omega _{\mu ab}$ are
the components of the spin connection, while $e$ is the inverse of the
vierbein determinant
\begin{equation}
e=\left( \det \left( e_{a}^{\;\;\mu }\right) \right) ^{-1}.  \label{xx2}
\end{equation}
In order to linearize action (\ref{xx1}), we develop the vierbein like
\begin{equation}
e_{a}^{\;\;\mu }=\delta _{a}^{\;\;\mu }-\frac{\lambda }{2}f_{a}^{\;\;\mu
},\;e=1+\frac{\lambda }{2}f,  \label{xx3}
\end{equation}
where $f$ is the trace of $f_{a}^{\;\;\mu }$. Consequently, we find that the
linearized form of (\ref{xx1}) reads as (we come back to the notations $\mu $%
, $\nu $, etc. for flat indices)
\begin{eqnarray}
S_{0}^{\prime }\left[ f_{\mu \nu },\omega _{\mu \alpha \beta }\right]
&=&\int d^{4}x\left( \omega _{\alpha }^{\;\;\alpha \mu }\left( \partial
_{\mu }f-\partial ^{\nu }f_{\mu \nu }\right) +\frac{1}{2}\omega ^{\mu \alpha
\beta }\partial _{\left[ \alpha \right. }f_{\left. \beta \right] \mu }\right.
\nonumber \\
&&\left. -\frac{1}{2}\left( \omega _{\alpha }^{\;\;\alpha \beta }\omega
_{\;\;\lambda \beta }^{\lambda }-\omega ^{\mu \alpha \beta }\omega _{\alpha
\mu \beta }\right) \right) .  \label{xx4}
\end{eqnarray}
We mention that the field $f_{\mu \nu }$ contains a symmetric, as well as an
antisymmetric part. The above linearized action is invariant under the gauge
transformations
\begin{equation}
\delta _{\epsilon }f_{\mu \nu }=\partial _{\mu }\epsilon _{\nu }+\epsilon
_{\mu \nu },\;\delta _{\epsilon }\omega _{\mu \alpha \beta }=\partial _{\mu
}\epsilon _{\alpha \beta ,}  \label{xy1}
\end{equation}
where the latter gauge parameters are antisymmetric, $\epsilon _{\alpha
\beta }=-\epsilon _{\beta \alpha }$. Eliminating the spin connection
components on their equations of motion (auxiliary fields) from (\ref{xx4})
\begin{equation}
\omega _{\mu \alpha \beta }\left( f\right) =\frac{1}{2}\left( \partial _{%
\left[ \mu \right. }f_{\left. \alpha \right] \beta }-\partial _{\left[ \mu
\right. }f_{\left. \beta \right] \alpha }-\partial _{\left[ \alpha \right.
}f_{\left. \beta \right] \mu }\right) ,  \label{xx5}
\end{equation}
we obtain the second-order action
\begin{eqnarray}
&&S_{0}^{\prime }\left[ f_{\mu \nu },\omega _{\mu \alpha \beta }\left(
f\right) \right] =S_{0}^{\prime \prime }\left[ f_{\mu \nu }\right] =-\int
d^{4}x\left( \frac{1}{8}\left( \partial ^{\left[ \mu \right. }f^{\left. \nu %
\right] \alpha }\right) \left( \partial _{\left[ \mu \right. }f_{\left. \nu %
\right] \alpha }\right) \right.  \nonumber \\
&&\left. +\frac{1}{4}\left( \partial ^{\left[ \mu \right. }f^{\left. \nu %
\right] \alpha }\right) \left( \partial _{\left[ \mu \right. }f_{\left.
\alpha \right] \nu }\right) -\frac{1}{2}\left( \partial _{\mu }f-\partial
^{\nu }f_{\mu \nu }\right) \left( \partial ^{\mu }f-\partial _{\alpha
}f^{\mu \alpha }\right) \right) ,  \label{xx6}
\end{eqnarray}
subject to the gauge invariances
\begin{equation}
\delta _{\epsilon }f_{\mu \nu }=\partial _{\left( \mu \right. }\epsilon
_{\left. \nu \right) }+\epsilon _{\mu \nu }.  \label{xx7}
\end{equation}
If we decompose $f_{\mu \nu }$ in its symmetric and antisymmetric parts
\begin{equation}
f_{\mu \nu }=h_{\mu \nu }+B_{\mu \nu },\;h_{\mu \nu }=h_{\nu \mu },\;B_{\mu
\nu }=-B_{\nu \mu },  \label{xx8}
\end{equation}
the action (\ref{xx6}) becomes
\begin{eqnarray}
S_{0}^{\prime \prime }\left[ f_{\mu \nu }\right] =S_{0}^{\prime \prime }%
\left[ h_{\mu \nu },B_{\mu \nu }\right] &=&\int d^{4}x\left( -\frac{1}{2}%
\left( \partial _{\mu }h_{\nu \rho }\right) \left( \partial ^{\mu }h^{\nu
\rho }\right) +\left( \partial _{\mu }h^{\mu \rho }\right) \left( \partial
^{\nu }h_{\nu \rho }\right) \right.  \nonumber \\
&&\left. \left( -\partial _{\mu }h\right) \left( \partial _{\nu }h^{\nu \mu
}\right) +\frac{1}{2}\left( \partial _{\mu }h\right) \left( \partial ^{\mu
}h\right) \right) ,  \label{xx9}
\end{eqnarray}
while the accompanying gauge transformations are given by
\begin{equation}
\delta _{\epsilon }h_{\mu \nu }=\partial _{\left( \mu \right. }\epsilon
_{\left. \nu \right) },\;\delta _{\epsilon }B_{\mu \nu }=\epsilon _{\mu \nu
}.  \label{xx10}
\end{equation}
It is easy to see that the right-hand side of (\ref{xx9}) is nothing but the
Pauli-Fierz action
\begin{equation}
S^{\prime \prime }\left[ h_{\mu \nu },B_{\mu \nu }\right] =S_{0}^{\left(
\mathrm{PF}\right) }\left[ h_{\mu \nu }\right] .  \label{xx11}
\end{equation}
Now, we show that the local BRST cohomologies associated with the
formulations (\ref{xx4})--(\ref{xy1}), (\ref{xx9})--(\ref{xx10}) and the
Pauli-Fierz model are isomorphic. As we have previously mentioned, we pass
from (\ref{xx4})--(\ref{xy1}) to (\ref{xx9})--(\ref{xx10}) via the
elimination of the auxiliary fields $\omega _{\mu \alpha \beta }$, such that
the general theorems from Section 15 of the first reference in~\cite{gen1}
ensure the isomorphism
\begin{equation}
H\left( s^{\prime }|d\right) \simeq H\left( s^{\prime \prime }|d\right) ,
\label{xx12}
\end{equation}
with $s^{\prime }$ and $s^{\prime \prime }$ the BRST differentials
corresponding to (\ref{xx4})--(\ref{xy1}) and respectively to (\ref{xx9})--(%
\ref{xx10}). On the other hand, we observe that the field $B_{\mu \nu }$
does not appear in (\ref{xx9}) and is subject to a shift gauge symmetry.
Thus, in any cohomological class from $H\left( s^{\prime \prime }|d\right) $
one can take a representative that is independent of $B_{\mu \nu }$, the
shift ghosts and all of their antifields. This is because these variables
form contractible pairs that drop out from $H\left( s^{\prime \prime
}|d\right) $ (see the general results from Section 14 of the first reference
in~\cite{gen1}). As a consequence, we have that
\begin{equation}
H\left( s^{\prime \prime }|d\right) \simeq H\left( s|d\right) ,  \label{xx13}
\end{equation}
where $s$ is the Pauli-Fierz BRST differential. Combining (\ref{xx12}) and (%
\ref{xx13}), we arrive at
\begin{equation}
H\left( s^{\prime }|d\right) \simeq H\left( s^{\prime \prime }|d\right)
\simeq H\left( s|d\right) .  \label{xx14}
\end{equation}
Because the local BRST cohomology (in ghost number equal to zero and one)
controls the deformation procedure, it results that the last isomorphisms
allow one to pass in a consistent manner from the Pauli-Fierz version to the
first- and second-order ones (in vierbein formulation) during the
deformation procedure.

\section{Analysis of the deformed theory\label{sec5}}

It is easy to see that one can go from (\ref{xx9})--(\ref{xx10}) to the
Pauli-Fierz version through the partial gauge-fixing $B_{\mu \nu }=0$. This
gauge-fixing is a consequence of the more general gauge-fixing condition~%
\cite{siegelfields}
\begin{equation}
\sigma _{\mu [a}e_{b]}^{\;\;\mu }=0.  \label{xx15}
\end{equation}
In the context of this partial gauge-fixing simple computation leads to the
vierbein fields and the inverse of their determinant up to the second order
in the coupling constant as
\begin{equation}
e_{a}^{\;\;\mu }=\stackrel{(0)}{e}_{a}^{\;\;\mu }+g\stackrel{(1)}{e}%
_{a}^{\;\;\mu }+g^{2}\stackrel{(2)}{e}_{a}^{\;\;\mu }+\cdots =\delta
_{a}^{\;\;\mu }-\frac{g}{2}h_{a}^{\;\;\mu
}+\frac{3g^{2}}{8}h_{a}^{\;\;\rho }h_{\rho }^{\;\;\mu }+\cdots ,
\label{PFD5.2}
\end{equation}
\begin{equation}
e=\stackrel{(0)}{e}+g\stackrel{(1)}{e}+g^{2}\stackrel{(2)}{e}+\cdots =1+\frac{g%
}{2}h+\frac{g^{2}}{8}\left( h^{2}-2h_{\mu \nu }h^{\mu \nu }\right) +\cdots .
\label{PFD5.3}
\end{equation}
Based on the isomorphisms (\ref{xx14}), we can further pass to the analysis
of the deformed theory obtained in the previous sections. The component of
antighost number equal to zero in $S_{1}^{\left( \mathrm{int}\right) }$ is
precisely the interacting lagrangian at order one in the coupling constant
\begin{eqnarray}
&&\mathcal{L}_{1}^{(\mathrm{int})}=a_{0}^{\left( \mathrm{int}\right)
}+a_{0}^{\left( \mathrm{Dirac}\right) }=\left[ \frac{1}{2}\bar{\psi}\left(
\mathrm{i}\gamma ^{\mu }\left( \partial _{\mu }\psi \right) -m\psi \right) h%
\right] +\left[ -\frac{\mathrm{i}}{2}\bar{\psi}\gamma ^{\alpha }\left(
\partial ^{\beta }\psi \right) h_{\alpha \beta }\right]  \nonumber \\
&&+\left[ -\frac{\mathrm{i}}{16}\bar{\psi}\gamma ^{\mu }\left[ \gamma
^{\alpha },\gamma ^{\beta }\right] \psi \partial _{[\alpha }h_{\beta ]\mu }%
\right] +\left[ M\left( \bar{\psi}\psi \right) \right]  \nonumber \\
&\equiv &\stackrel{(1)}{e}\mathcal{L}_{0}^{(\mathrm{D})}+\stackrel{(0)}{e}%
\stackrel{(1)}{e}_{a}^{\;\;\mu }\bar{\psi}\mathrm{i}\gamma ^{a}\stackrel{(0)}{D%
}_{\mu }\psi +\stackrel{(0)}{e}\stackrel{(0)}{e}_{a}^{\;\;\mu }\bar{\psi}%
\mathrm{i}\gamma ^{a}\stackrel{(1)}{D}_{\mu }\psi
+\stackrel{(0)}{e}M\left( \bar{\psi}\psi \right) ,  \label{id1}
\end{eqnarray}
where
\begin{equation}
\stackrel{(0)}{D}_{\mu }=\partial _{\mu },  \label{xx19}
\end{equation}
and
\begin{equation}
\stackrel{(1)}{D}_{\mu }=\frac{1}{16}\stackrel{\left( 1\right)
}{\omega }_{\mu ab}\left[ \gamma ^{a},\gamma ^{b}\right] ,
\label{uv1}
\end{equation}
with
\begin{equation}
\stackrel{\left( 1\right) }{\omega }_{\mu ab}=-\partial
_{[a}h_{b]\mu } \label{uv4}
\end{equation}
the linearized form of the full spin-connection
\begin{eqnarray}
\omega _{\mu ab} &=&e_{b}^{\;\;\nu }\partial _{\nu }e_{a\mu }-e_{a}^{\;\;\nu
}\partial _{\nu }e_{b\mu }+e_{a\nu }\partial _{\mu }e_{b}^{\;\;\nu }
\nonumber \\
&&-e_{b\nu }\partial _{\mu }e_{a}^{\;\;\nu }+e_{\left[ a\right. }^{\;\;\rho
}e_{\left. b\right] }^{\;\;\nu }e_{c\mu }\partial _{\nu }e_{\;\;\rho }^{c}
\nonumber \\
&=&g\stackrel{\left( 1\right) }{\omega }_{\mu
ab}+g^{2}\stackrel{\left( 2\right) }{\omega }_{\mu ab}+\cdots .
\label{uv2}
\end{eqnarray}
In (\ref{uv2}) $e_{a\mu }$ represents the inverse of the vierbein field.
Along the same line, the piece of antighost number equal to zero from the
second-order deformation offers us the interacting lagrangian at order two
in the coupling constant
\begin{eqnarray}
&&\mathcal{L}_{2}^{(\mathrm{int})}=b_{0}^{\left( \mathrm{int}\right) }=\left[
\frac{1}{8}\left( \bar{\psi}\mathrm{i}\gamma ^{\mu }\left( \partial _{\mu
}\psi \right) -m\bar{\psi}\psi \right) \left( h^{2}-2h_{\mu \nu }h^{\mu \nu
}\right) \right] +\left[ \frac{1}{2}hM\left( \bar{\psi}\psi \right) \right]
\nonumber \\
&&+\left[ -\frac{\mathrm{i}}{4}\bar{\psi}\gamma ^{\mu }\left( \partial ^{\nu
}\psi \right) hh_{\mu \nu }\right] +\left[ -\frac{\mathrm{i}}{32}\bar{\psi}%
\gamma ^{\mu }\left[ \gamma ^{\alpha },\gamma ^{\beta }\right] \psi
h\partial _{[\alpha }h_{\beta ]\mu }\right]  \nonumber \\
&&+\left[ \frac{3\mathrm{i}}{8}\bar{\psi}\gamma ^{\mu }\left( \partial ^{\nu
}\psi \right) h_{\mu \sigma }h_{\nu }^{\sigma }\right] +\left[ \frac{\mathrm{%
i}}{32}\bar{\psi}\gamma ^{\mu }\left[ \gamma ^{\alpha },\gamma ^{\beta }%
\right] \psi h_{\mu }^{\sigma }\partial _{[\alpha }h_{\beta ]\sigma }\right]
\nonumber \\
&&+\left[ -\frac{\mathrm{i}}{64}\bar{\psi}\gamma ^{\mu }\left[ \gamma
^{\alpha },\gamma ^{\beta }\right] \psi \left( 2h_{[\alpha }^{\rho }\left(
\partial _{\beta ]}h_{\rho \mu }\right) +2\left( \partial _{\rho }h_{\mu
[\alpha }\right) h_{\beta ]}^{\rho }-\left( \partial _{\mu }h_{\rho [\alpha
}\right) h_{\beta ]}^{\rho }\right) \right]  \nonumber \\
&\equiv &\stackrel{(2)}{e}\mathcal{L}_{0}^{(\mathrm{D})}+\stackrel{(1)}{e}%
M\left( \bar{\psi}\psi \right) +\stackrel{(1)}{e}\stackrel{(1)}{e}%
_{a}^{\;\;\mu }\bar{\psi}\mathrm{i}\gamma ^{a}\stackrel{(0)}{D}_{\mu }\psi +%
\stackrel{(1)}{e}\stackrel{(0)}{e}_{a}^{\;\;\mu
}\bar{\psi}\mathrm{i}\gamma
^{a}\stackrel{(1)}{D}_{\mu }\psi  \nonumber \\
&&+\stackrel{(0)}{e}\stackrel{(2)}{e}_{a}^{\;\;\mu
}\bar{\psi}\mathrm{i}\gamma
^{a}\stackrel{(0)}{D}_{\mu }\psi +\stackrel{(0)}{e}\stackrel{(1)}{e}%
_{a}^{\;\;\mu }\bar{\psi}\mathrm{i}\gamma ^{a}\stackrel{(1)}{D}_{\mu }\psi +%
\stackrel{(0)}{e}\stackrel{(0)}{e}_{a}^{\;\;\mu
}\bar{\psi}\mathrm{i}\gamma ^{a}\stackrel{(2)}{D}_{\mu }\psi ,
\label{id2}
\end{eqnarray}
where
\begin{equation}
\stackrel{(2)}{D}_{\mu }=\frac{1}{16}\stackrel{\left( 2\right)
}{\omega }_{\mu ab}\left[ \gamma ^{a},\gamma ^{b}\right] ,
\label{uv6}
\end{equation}
while
\begin{equation}
\stackrel{\left( 2\right) }{\omega }_{\mu ab}=-\frac{1}{4}\left(
2h_{c[a}\left( \partial _{b]}h_{\;\;\mu }^{c}\right) -2h_{\left[
a\right. }^{\;\;\;\nu }\partial _{\nu }h_{\left. b\right] \mu
}-\left( \partial _{\mu }h_{[a}^{\;\;\;\nu }\right) h_{b]\nu
}\right)  \label{uv5}
\end{equation}
is the second-order approximation of the spin-connection. With the help of (%
\ref{id1}) and (\ref{id2}) we deduce that $\mathcal{L}_{0}^{(\mathrm{D})}+g%
\mathcal{L}_{1}^{(\mathrm{int})}+g^{2}\mathcal{L}_{2}^{(\mathrm{int}%
)}+\cdots $ comes from expanding the fully deformed lagrangian
\begin{equation}
\mathcal{L}^{\left( \mathrm{int}\right) }=e\bar{\psi}\left( \mathrm{i}%
e_{a}^{\;\;\mu }\gamma ^{a}D_{\mu }\psi -m\psi \right) +geM\left( \bar{\psi}%
\psi \right) ,  \label{PFD5.1}
\end{equation}
where
\begin{equation}
D_{\mu }\psi =\partial _{\mu }\psi +\frac{1}{16}\omega _{\mu ab}\left[
\gamma ^{a},\gamma ^{b}\right] \psi  \label{xx20}
\end{equation}
is the full covariant derivative of $\psi $.

The pieces linear in the antifields $\psi ^{*}$ and $\bar{\psi}^{*}$ from
the deformed solution to the master equation give us the deformed gauge
transformations for the Dirac fields as
\begin{eqnarray}
\delta _{\epsilon }\psi &=&g\left( \partial ^{\alpha }\psi \right) \epsilon
_{\alpha }+\frac{g}{16}\left[ \gamma ^{\alpha },\gamma ^{\beta }\right] \psi
\partial _{[\alpha }\epsilon _{\beta ]}-\frac{g^{2}}{2}\left( \partial
^{\alpha }\psi \right) \epsilon ^{\beta }h_{\alpha \beta }  \nonumber \\
&&-\frac{g^{2}}{32}\left[ \gamma ^{\alpha },\gamma ^{\beta }\right] \psi
\left( \epsilon ^{\sigma }\partial _{[\alpha }h_{\beta ]\sigma }-\frac{1}{2}%
h_{\sigma [\alpha }\left( \partial _{\beta ]}\epsilon ^{\sigma }-\partial
^{\sigma }\epsilon _{\beta ]}\right) \right) +\cdots  \nonumber \\
&=&g\stackrel{(1)}{\delta }_{\epsilon }\psi +g^{2}\stackrel{(2)}{\delta }%
_{\epsilon }\psi +\cdots ,  \label{PFD5.5}
\end{eqnarray}
\begin{eqnarray}
\delta _{\epsilon }\bar{\psi} &=&g\left( \partial ^{\alpha }\bar{\psi}%
\right) \epsilon _{\alpha }-\frac{g}{16}\bar{\psi}\left[ \gamma ^{\alpha
},\gamma ^{\beta }\right] \partial _{[\alpha }\epsilon _{\beta ]}-\frac{g^{2}%
}{2}\left( \partial ^{\alpha }\bar{\psi}\right) \epsilon ^{\beta }h_{\alpha
\beta }  \nonumber \\
&&+\frac{g^{2}}{32}\bar{\psi}\left[ \gamma ^{\alpha },\gamma ^{\beta }\right]
\left( \epsilon ^{\sigma }\partial _{[\alpha }h_{\beta ]\sigma }-\frac{1}{2}%
h_{\sigma [\alpha }\left( \partial _{\beta ]}\epsilon ^{\sigma }-\partial
^{\sigma }\epsilon _{\beta ]}\right) \right) +\cdots  \nonumber \\
&=&g\stackrel{(1)}{\delta }_{\epsilon }\bar{\psi}+g^{2}\stackrel{(2)}{\delta }%
_{\epsilon }\bar{\psi}+\cdots .  \label{PFD5.6}
\end{eqnarray}
The first two orders of the gauge transformations can be put under the form
\begin{eqnarray}
\stackrel{(1)}{\delta }_{\epsilon }\psi &=&\left( \partial _{\mu
}\psi \right) \stackrel{(0)}{\bar{\epsilon}}^{\mu
}+\frac{1}{8}\left[ \gamma
^{a},\gamma ^{b}\right] \psi \stackrel{(0)}{\epsilon }_{ab},  \label{uw1} \\
\stackrel{(2)}{\delta }_{\epsilon }\psi &=&\left( \partial _{\mu
}\psi \right) \stackrel{(1)}{\bar{\epsilon}}^{\mu
}+\frac{1}{8}\left[ \gamma
^{a},\gamma ^{b}\right] \psi \stackrel{(1)}{\epsilon }_{ab},  \label{uw2} \\
\stackrel{(1)}{\delta }_{\epsilon }\bar{\psi} &=&\left( \partial _{\mu }\bar{%
\psi}\right) \stackrel{(0)}{\bar{\epsilon}}^{\mu
}-\frac{1}{8}\bar{\psi}\left[
\gamma ^{a},\gamma ^{b}\right] \stackrel{(0)}{\epsilon }_{ab},  \label{uw3} \\
\stackrel{(2)}{\delta }_{\epsilon }\bar{\psi} &=&\left( \partial _{\mu }\bar{%
\psi}\right) \stackrel{(1)}{\bar{\epsilon}}^{\mu
}-\frac{1}{8}\bar{\psi}\left[ \gamma ^{a},\gamma ^{b}\right]
\stackrel{(1)}{\epsilon }_{ab},  \label{uw4}
\end{eqnarray}
where we used the notations
\begin{eqnarray}
\stackrel{(0)}{\bar{\epsilon}}^{\mu } &=&\epsilon ^{\mu }=\epsilon
^{a}\delta _{a}^{\;\;\mu },\;\stackrel{(1)}{\bar{\epsilon}}^{\mu
}=-\frac{1}{2}\epsilon
^{a}h_{a}^{\;\;\mu },  \label{uv16} \\
\stackrel{(0)}{\epsilon }_{ab} &=&\frac{1}{2}\partial _{[a}\epsilon
_{b]},
\label{apx0} \\
\stackrel{(1)}{\epsilon }_{ab} &=&-\frac{1}{4}\epsilon ^{c}\partial
_{[a}h_{b]c}+\frac{1}{8}h_{[a}^{c}\partial _{b]}^{\left. {}\right.
}\epsilon _{c}+\frac{1}{8}\left( \partial _{c}\epsilon _{[a}^{\left.
{}\right. }\right) h_{b]}^{c}.  \label{apx1}
\end{eqnarray}
Based on these notations, the gauge transformations of the spinors take the
form
\begin{eqnarray}
\delta _{\epsilon }\psi &=&g\left( \left( \partial _{\mu }\psi \right)
\left( \stackrel{(0)}{\bar{\epsilon}}^{\mu }+g\stackrel{(1)}{\bar{\epsilon}}%
^{\mu }+\cdots \right) \right.  \nonumber \\
&&\left. +\frac{1}{8}\left[ \gamma ^{a},\gamma ^{b}\right] \psi
\left( \stackrel{(0)}{\epsilon }_{ab}+g\stackrel{(1)}{\epsilon
}_{ab}+\cdots \right) \right) ,  \label{uw5}
\end{eqnarray}
\begin{eqnarray}
\delta _{\epsilon }\bar{\psi} &=&g\left( \left( \partial _{\mu }\bar{\psi}%
\right) \left( \stackrel{(0)}{\bar{\epsilon}}^{\mu }+g\stackrel{(1)}{\bar{%
\epsilon}}^{\mu }+\cdots \right) \right.  \nonumber \\
&&\left. -\frac{1}{8}\bar{\psi}\left[ \gamma ^{a},\gamma ^{b}\right]
\left( \stackrel{(0)}{\epsilon }_{ab}+g\stackrel{(1)}{\epsilon
}_{ab}+\cdots \right) \right) .  \label{uw6}
\end{eqnarray}
The gauge parameters $\stackrel{(0)}{\epsilon }_{ab}$ si $\stackrel{(1)}{%
\epsilon }_{ab}$ are precisely the first two terms from the Lorentz
parameters expressed in terms of the flat parameters $\epsilon ^{a}$ via the
partial gauge-fixing (\ref{xx15}). Indeed, (\ref{xx15}) leads to
\begin{equation}
\delta _{\epsilon }\left( \sigma _{\mu [a}e_{b]}^{\;\;\mu }\right) =0,
\label{id5}
\end{equation}
where
\begin{equation}
\delta _{\epsilon }e_{a}^{\;\;\mu }=\bar{\epsilon}^{\rho }\partial _{\rho
}e_{a}^{\;\;\mu }-e_{a}^{\;\;\rho }\partial _{\rho }\bar{\epsilon}^{\mu
}+\epsilon _{a}^{\;\;b}e_{b}^{\;\;\mu }.  \label{id6}
\end{equation}
Substituting (\ref{PFD5.2}) together with the expansions
\begin{equation}
\bar{\epsilon}^{\mu }=\stackrel{(0)}{\bar{\epsilon}}^{\mu }+g\stackrel{(1)}{%
\bar{\epsilon}}^{\mu }+\cdots =\left( \delta _{a}^{\;\;\mu }-\frac{g}{2}%
h_{a}^{\;\;\mu }+\cdots \right) \epsilon ^{a}  \label{uv15}
\end{equation}
and
\begin{equation}
\epsilon _{ab}=\stackrel{(0)}{\epsilon }_{ab}+g\stackrel{(1)}{\epsilon }%
_{ab}+\cdots  \label{uv12}
\end{equation}
in (\ref{id5}), we arrive precisely to (\ref{apx0})--(\ref{apx1}). At this
point it is easy to see that the gauge transformations (\ref{uw5})--(\ref%
{uw6}) come from the perturbative expansion of the full gauge
transformations
\begin{equation}
\delta _{\epsilon }\psi =g\left( \left( \partial _{\mu }\psi \right) \bar{%
\epsilon}^{\mu }+\frac{1}{8}\left[ \gamma ^{a},\gamma ^{b}\right] \psi
\epsilon _{ab}\right) ,  \label{full}
\end{equation}
\begin{equation}
\delta _{\epsilon }\bar{\psi}=g\left( \left( \partial _{\mu }\bar{\psi}%
\right) \bar{\epsilon}^{\mu }-\frac{1}{8}\bar{\psi}\left[ \gamma ^{a},\gamma
^{b}\right] \epsilon _{ab}\right) .  \label{uw7}
\end{equation}
The full gauge transformations can be suggestively written like
\begin{equation}
\delta _{\epsilon }\psi =g\left( \left( \partial _{\mu }\psi \right) \bar{%
\epsilon}^{\mu }+\frac{1}{2}\Sigma ^{ab}\psi \epsilon _{ab}\right) ,
\label{full1}
\end{equation}
\begin{equation}
\delta _{\epsilon }\bar{\psi}=g\left( \left( \partial _{\mu }\bar{\psi}%
\right) \bar{\epsilon}^{\mu }-\frac{1}{2}\bar{\psi}\Sigma ^{ab}\epsilon
_{ab}\right) ,  \label{uw8}
\end{equation}
where
\begin{equation}
\Sigma ^{ab}=\frac{1}{4}\left[ \gamma ^{a},\gamma ^{b}\right]  \label{uw9}
\end{equation}
are the spin operators, whose commutators read as
\begin{equation}
\left[ \Sigma ^{ab},\Sigma ^{cd}\right] =\sigma ^{a[c}\Sigma ^{d]b}-\sigma
^{b[c}\Sigma ^{d]a}.  \label{uw10}
\end{equation}
In conclusion, the interaction between a Dirac field and one spin-two field
leads to the interacting lagrangian (\ref{PFD5.1}), while the gauge
transformations of the Dirac spinors are given by (\ref{full1}) and (\ref%
{uw8}).

\section{Impossibility of cross-interactions between gravitons in the
presence of the Dirac field}

As it has been proved in~\cite{multi}, there are no direct cross-couplings
that can be introduced in a finite collection of gravitons and also no
intermediate cross-couplings between different gravitons in the presence of
a scalar field. In this section, under the hypotheses of locality,
smoothness of the interactions in the coupling constant, Poincar\'{e}
invariance, (background) Lorentz invariance and the preservation of the
number of derivatives on each field, we will prove that there are no
intermediate cross-couplings between different gravitons in the presence of
a Dirac field.

Now, we start from a sum of Pauli-Fierz actions and a Dirac action
\begin{eqnarray}
S_{0}^{\mathrm{L}}\left[ h_{\mu \nu }^{A},\psi ,\bar{\psi}\right] &=&\int
d^{4}x\left( -\frac{1}{2}\left( \partial _{\mu }h_{\nu \rho }^{A}\right)
\left( \partial ^{\mu }h_{A}^{\nu \rho }\right) +\left( \partial _{\mu
}h_{A}^{\mu \rho }\right) \left( \partial ^{\nu }h_{\nu \rho }^{A}\right)
\right.  \nonumber \\
&&\left. -\left( \partial _{\mu }h^{A}\right) \left( \partial _{\nu
}h_{A}^{\nu \mu }\right) +\frac{1}{2}\left( \partial _{\mu }h^{A}\right)
\left( \partial ^{\mu }h_{A}\right) \right)  \nonumber \\
&&+\int d^{4}x\bar{\psi}\left( \mathrm{i}\gamma ^{\mu }\left( \partial _{\mu
}\psi \right) -m\psi \right) ,  \label{vu1}
\end{eqnarray}
where $h_{A}$ is the trace of the field $h_{A}^{\mu \nu }$ ($h_{A}=\sigma
_{\mu \nu }h_{A}^{\mu \nu }$), while $A=1,2,\cdots ,n$. The gauge
transformations of the action (\ref{vu1}) are
\begin{equation}
\delta _{\epsilon }h_{\mu \nu }^{A}=\partial _{(\mu }\epsilon _{\nu
)}^{A},\;\delta _{\epsilon }\psi =\delta _{\epsilon }\bar{\psi}=0.  \label{5}
\end{equation}
The BRST complex comprises the fields/ghosts
\begin{equation}
\phi ^{\alpha _{0}}=\left( h_{\mu \nu }^{A},\psi ,\bar{\psi}\right) ,\;\eta
_{\mu }^{A},  \label{vu2}
\end{equation}
and respectively the antifields
\begin{equation}
\phi _{\alpha _{0}}^{*}=\left( h_{A}^{*\mu \nu },\psi ^{*},\bar{\psi}%
^{*}\right) ,\;\eta _{A}^{*\mu }.  \label{vu3}
\end{equation}
The BRST differential splits in this situation like in (\ref{PFD7}), while
the actions of $\delta $ and $\gamma $ on the BRST generators are defined by
\begin{eqnarray}
\delta h_{A}^{*\mu \nu } &=&2H_{A}^{\mu \nu },\;\delta \psi ^{*}=-\left( m%
\bar{\psi}+\mathrm{i}\partial _{\mu }\bar{\psi}\gamma ^{\mu }\right) ,
\label{12} \\
\delta \bar{\psi}^{*} &=&-\left( \mathrm{i}\gamma ^{\mu }\partial _{\mu
}\psi -m\psi \right) ,\;\delta \eta _{A}^{*\mu }=-2\partial _{\nu
}h_{A}^{*\mu \nu },  \label{13} \\
\delta \phi ^{\alpha _{0}} &=&0,\;\delta \eta _{\mu }^{A}=0,  \label{14} \\
\gamma \phi _{\alpha _{0}}^{*} &=&0,\;\gamma \eta _{A}^{*\mu }=0,  \label{15}
\\
\gamma h_{\mu \nu }^{A} &=&\partial _{(\mu }\eta _{\nu )}^{A},\;\gamma \psi
=\gamma \bar{\psi}=0,\;\gamma \eta _{\mu }^{A}=0,  \label{16}
\end{eqnarray}
where $H_{A}^{\mu \nu }=K_{A}^{\mu \nu }-\frac{1}{2}\sigma ^{\mu \nu }K_{A}$
is the linearized Einstein tensor for the field $h_{A}^{\mu \nu }$. In this
case the solution to the master equation reads as
\begin{equation}
\bar{S}=S_{0}^{\mathrm{L}}\left[ h_{\mu \nu }^{A},\psi ,\bar{\psi}\right]
+\int d^{4}x\left( h_{A}^{*\mu \nu }\partial _{(\mu }\eta _{\nu
)}^{A}\right) .  \label{18}
\end{equation}

The first-order deformation of the solution to the master equation may be
decomposed in a manner similar to the case of a single graviton
\begin{equation}
\alpha =\alpha ^{\left( \mathrm{PF}\right) }+\alpha ^{\left( \mathrm{int}%
\right) }+\alpha ^{\left( \mathrm{Dirac}\right) }.  \label{vux}
\end{equation}
The first-order deformation in the Pauli-Fierz sector, $\alpha ^{\left(
\mathrm{PF}\right) }$, is of the form~\cite{multi}
\begin{equation}
\alpha ^{\left( \mathrm{PF}\right) }=\alpha _{2}^{\left( \mathrm{PF}\right)
}+\alpha _{1}^{\left( \mathrm{PF}\right) }+\alpha _{0}^{\left( \mathrm{PF}%
\right) },  \label{vx1}
\end{equation}
with
\begin{equation}
\alpha _{2}^{\left( \mathrm{PF}\right) }=\frac{1}{2}f_{BC}^{A}\eta
_{A}^{*\mu }\eta ^{B\nu }\partial _{[\mu }^{\left. {}\right. }\eta _{\nu
]}^{C}.  \label{vx2}
\end{equation}
In (\ref{vx2}), all the coefficients $f_{BC}^{A}$ are constant. The
condition that $\alpha _{2}^{\left( \mathrm{PF}\right) }$ indeed produces a
consistent $\alpha _{1}^{\left( \mathrm{PF}\right) }$ implies that these
constants must be symmetric in their lower indices~\cite{multi}\footnote{%
The term (\ref{vx2}) differs from that corresponding to~\cite{multi} through
a $\gamma $-exact term, which does not affect (\ref{vx3}).}
\begin{equation}
f_{BC}^{A}=f_{CB}^{A}.  \label{vx3}
\end{equation}
With (\ref{vx3}) at hand, we find that
\begin{equation}
\alpha _{1}^{\left( \mathrm{PF}\right) }=f_{BC}^{A}h_{A}^{*\mu \rho }\left(
\left( \partial _{\rho }\eta ^{B\nu }\right) h_{\mu \nu }^{C}-\eta ^{B\nu
}\partial _{[\mu }^{\left. {}\right. }h_{\nu ]\rho }^{C}\right) .
\label{vxx}
\end{equation}
The requirement that $\alpha _{1}^{\left( \mathrm{PF}\right) }$ leads to a
consistent $\alpha _{0}^{\left( \mathrm{PF}\right) }$ implies that~\cite%
{multi}\footnote{%
The piece (\ref{vxx}) differs from that corresponding to~\cite{multi}
through a $\delta $-exact term, which does not change (\ref{3.47}).}
\begin{equation}
f_{ABC}=\frac{1}{3}f_{\left( ABC\right) },  \label{3.47}
\end{equation}
where, by definition, $f_{ABC}=\delta _{AD}f_{BC}^{D}$. Based on (\ref{3.47}%
), we obtain that the resulting $\alpha _{0}^{\left( \mathrm{PF}\right) }$
reads as in~\cite{multi} (where this component is denoted by $a_{0}$ and $%
f_{ABC}$ by $a_{abc}$).

If we go along exactly the same line like in the subsection 4.2, we get that
$\alpha ^{\left( \mathrm{int}\right) }=\alpha _{1}^{\left( \mathrm{int}%
\right) }+\alpha _{0}^{\left( \mathrm{int}\right) }$, where
\begin{eqnarray}
\alpha _{1}^{\left( \mathrm{int}\right) } &=&k_{A}\left( \psi ^{*}\left(
\partial ^{\alpha }\psi \right) +\left( \partial ^{\alpha }\bar{\psi}\right)
\bar{\psi}^{*}\right) \eta _{\alpha }^{A}  \nonumber \\
&&-\frac{k_{A}}{16}\left( \bar{\psi}\left[ \gamma ^{\alpha },\gamma ^{\beta }%
\right] \bar{\psi}^{*}-\psi ^{*}\left[ \gamma ^{\alpha },\gamma ^{\beta }%
\right] \psi \right) \partial _{[\alpha }\eta _{\beta ]}^{A},  \label{3.35}
\end{eqnarray}
\begin{eqnarray}
\alpha _{0}^{\left( \mathrm{int}\right) } &=&\frac{k_{A}}{2}\left( \bar{\psi}%
\left( \mathrm{i}\gamma ^{\mu }\left( \partial _{\mu }\psi \right) -m\psi
\right) h^{A}-\mathrm{i}\bar{\psi}\gamma ^{\alpha }\left( \partial ^{\beta
}\psi \right) h_{\alpha \beta }^{A}\right)  \nonumber \\
&&-\frac{\mathrm{i}k_{A}}{16}\bar{\psi}\gamma ^{\mu }\left[ \gamma ^{\alpha
},\gamma ^{\beta }\right] \psi \partial _{[\alpha }h_{\beta ]\mu }^{A},
\label{3.36}
\end{eqnarray}
and $k_{A}$ are some real constants. Meanwhile, we find in a direct manner
that
\begin{equation}
\alpha ^{\left( \mathrm{Dirac}\right) }=a_{0}^{\left( \mathrm{Dirac}\right)
},  \label{vuw}
\end{equation}
with $a_{0}^{\left( \mathrm{Dirac}\right) }$ given in (\ref{PFD3.50}).

Let us investigate next the consistency of the first-order deformation. If
we perform the notations
\begin{eqnarray}
\hat{S}_{1}^{\left( \mathrm{PF}\right) } &=&\int d^{4}x\alpha ^{\left(
\mathrm{PF}\right) },  \label{vx4} \\
\hat{S}_{1}^{\left( \mathrm{int}\right) } &=&\int d^{4}x\left( \alpha
^{\left( \mathrm{int}\right) }+\alpha ^{\left( \mathrm{Dirac}\right)
}\right) ,  \label{4.1} \\
\hat{S}_{1} &=&\hat{S}_{1}^{\left( \mathrm{PF}\right) }+\hat{S}_{1}^{\left(
\mathrm{int}\right) },  \label{vx5}
\end{eqnarray}%
then the equation $\left( \hat{S}_{1},\hat{S}_{1}\right)
+2s\hat{S}_{2}=0$ (expressing the consistency of the first-order
deformation) equivalently splits into the equations
\begin{eqnarray}
\left( \hat{S}_{1}^{\left( \mathrm{PF}\right) },\hat{S}_{1}^{\left( \mathrm{%
PF}\right) }\right) +2s\hat{S}_{2}^{\left( \mathrm{PF}\right) } &=&0,
\label{vx6} \\
2\left( \hat{S}_{1}^{\left( \mathrm{PF}\right) },\hat{S}_{1}^{\left( \mathrm{%
int}\right) }\right) +\left( \hat{S}_{1}^{\left( \mathrm{int}\right) },\hat{S%
}_{1}^{\left( \mathrm{int}\right) }\right) +2s\hat{S}_{2}^{\left( \mathrm{int%
}\right) } &=&0,  \label{vx7}
\end{eqnarray}%
where $\hat{S}_{2}=\hat{S}_{2}^{\left( \mathrm{PF}\right) }+\hat{S}%
_{2}^{\left( \mathrm{int}\right) }$. The equation (\ref{vx6}) requires that
the constants $f_{AB}^{C}$ satisfy the supplementary conditions~\cite{multi}
\begin{equation}
f_{A[B}^{D}f_{C]D}^{E}=0,  \label{4.20}
\end{equation}%
so they are the structure constants of a finite-dimensional, commutative,
symmetric and associative real algebra $\mathcal{A}$. The analysis realized
in~\cite{multi} shows us that such an algebras has a trivial structure
(being expressed like a direct sum of some one-dimensional ideals). So we
obtain that
\begin{equation}
f_{AB}^{C}=0\quad \mathrm{if}\quad A\neq B.  \label{4.21}
\end{equation}

Let us analyze now the equation (\ref{vx7}). If we denote by $\hat{\Delta}%
^{\left( \mathrm{int}\right) }$ and $\beta ^{\left( \mathrm{int}\right) }$
the non-integrated densities of the functionals $2\left( \hat{S}_{1}^{\left(
\mathrm{PF}\right) },\hat{S}_{1}^{\left( \mathrm{int}\right) }\right)
+\left( \hat{S}_{1}^{\left( \mathrm{int}\right) },\hat{S}_{1}^{\left(
\mathrm{int}\right) }\right) $ and respectively of $\hat{S}_{2}^{\left(
\mathrm{int}\right) }$, then the equation (\ref{vx7}) in local form becomes
\begin{equation}
\hat{\Delta}^{\left( \mathrm{int}\right) }=-2s\beta ^{\left( \mathrm{int}%
\right) }+\partial _{\mu }k^{\mu },  \label{4.5}
\end{equation}
with
\begin{equation}
\mathrm{gh}\left( \hat{\Delta}^{\left( \mathrm{int}\right) }\right) =1,\;%
\mathrm{gh}\left( \beta ^{\left( \mathrm{int}\right) }\right) =0,\;\mathrm{gh%
}\left( k^{\mu }\right) =1.  \label{45a}
\end{equation}
The computation of $\hat{\Delta}^{\left( \mathrm{int}\right) }$ reveals in
our case the following decomposition along the antighost number
\begin{equation}
\hat{\Delta}^{\left( \mathrm{int}\right) }=\hat{\Delta}_{0}^{\left( \mathrm{%
int}\right) }+\hat{\Delta}_{1}^{\left( \mathrm{int}\right) },\;\mathrm{agh}%
\left( \hat{\Delta}_{I}^{\left( \mathrm{int}\right) }\right) =I,\;I=0,1,
\label{4.6}
\end{equation}
with
\begin{eqnarray}
\hat{\Delta}_{1}^{\left( \mathrm{int}\right) } &=&\gamma \left(
k_{A}k_{B}\left( \psi ^{*}\left( \partial ^{\alpha }\psi \right) +\left(
\partial ^{\alpha }\bar{\psi}\right) \bar{\psi}^{*}\right) \eta ^{A\beta
}h_{\alpha \beta }^{B}\right.  \nonumber \\
&&\left. -\frac{1}{16}M^{\alpha \beta }\left( \left(
2k_{A}k_{B}-k_{D}f_{AB}^{D}\right) \eta ^{A\sigma }\partial _{[\alpha
}h_{\beta ]\sigma }^{B}-k_{D}f_{AB}^{D}h_{\alpha }^{A\rho }\partial _{[\beta
}^{\left. {}\right. }\eta _{\rho ]}^{B}\right) \right)  \nonumber \\
&&+\left( k_{D}f_{AB}^{D}-k_{A}k_{B}\right) \left( \left( \psi ^{*}\left(
\partial ^{\alpha }\psi \right) +\left( \partial ^{\alpha }\bar{\psi}\right)
\bar{\psi}^{*}\right) \eta ^{A\beta }\partial _{[\alpha }\eta _{\beta
]}^{B}\right.  \nonumber \\
&&\left. -\frac{1}{16}M^{\alpha \beta }\sigma ^{\mu \nu }\partial _{[\alpha
}\eta _{\mu ]}^{A}\partial _{[\beta }\eta _{\nu ]}^{B}\right) ,
\label{4.12a}
\end{eqnarray}
where we used the notation
\begin{equation}
M^{\alpha \beta }=\bar{\psi}\left[ \gamma ^{\alpha },\gamma ^{\beta }\right]
\bar{\psi}^{*}-\psi ^{*}\left[ \gamma ^{\alpha },\gamma ^{\beta }\right]
\psi .  \label{4.12b}
\end{equation}
The concrete form of $\hat{\Delta}_{0}^{\left( \mathrm{int}\right) }$ is not
important in what follows and therefore we will skip it. Due to the
decomposition (\ref{4.6}), we have that $\beta ^{\left( \mathrm{int}\right)
} $ and $k^{\mu }$ from (\ref{4.5}) can be decomposed like
\begin{eqnarray}
\beta ^{\left( \mathrm{int}\right) } &=&\beta _{0}^{\left( \mathrm{int}%
\right) }+\beta _{1}^{\left( \mathrm{int}\right) }+\beta _{2}^{\left(
\mathrm{int}\right) },\;\mathrm{agh}\left( \beta _{I}^{\left( \mathrm{int}%
\right) }\right) =I,\;I=0,1,2,  \label{4.7} \\
k^{\mu } &=&k_{0}^{\mu }+k_{1}^{\mu }+k_{2}^{\mu },\;\mathrm{agh}\left(
k_{I}^{\mu }\right) =I,\;I=0,1,2.  \label{4.8}
\end{eqnarray}
By projecting the equation (\ref{4.5}) on various values of the antighost
number, we obtain the tower of equations
\begin{eqnarray}
\gamma \beta _{2}^{\left( \mathrm{int}\right) } &=&\partial _{\mu }\left(
\frac{1}{2}k_{2}^{\mu }\right) ,  \label{4.9a} \\
\hat{\Delta}_{1}^{\left( \mathrm{int}\right) } &=&-2\left( \delta \beta
_{2}^{\left( \mathrm{int}\right) }+\gamma \beta _{1}^{\left( \mathrm{int}%
\right) }\right) +\partial _{\mu }k_{1}^{\mu },  \label{4.9} \\
\hat{\Delta}_{0}^{\left( \mathrm{int}\right) } &=&-2\left( \delta \beta
_{1}^{\left( \mathrm{int}\right) }+\gamma \beta _{0}^{\left( \mathrm{int}%
\right) }\right) +\partial _{\mu }k_{0}^{\mu }.  \label{4.10}
\end{eqnarray}
By a trivial redefinition, the equation (\ref{4.9a}) can always be replaced
with
\begin{equation}
\gamma \beta _{2}^{\left( \mathrm{int}\right) }=0.  \label{4.10a}
\end{equation}
Analyzing the expression of $\hat{\Delta}_{1}^{\left( \mathrm{int}\right) }$
in (\ref{4.12a}) we observe that it can be expressed as in (\ref{4.9}) if
\begin{eqnarray}
\hat{\chi} &=&\left( k_{D}f_{AB}^{D}-k_{A}k_{B}\right) \left( \left( \psi
^{*}\left( \partial ^{\alpha }\psi \right) +\left( \partial ^{\alpha }\bar{%
\psi}\right) \bar{\psi}^{*}\right) \eta ^{A\beta }\partial _{[\alpha }\eta
_{\beta ]}^{B}\right.  \nonumber \\
&&\left. -\frac{1}{16}M^{\alpha \beta }\sigma ^{\mu \nu }\partial _{[\alpha
}\eta _{\mu ]}^{A}\partial _{[\beta }\eta _{\nu ]}^{B}\right) ,
\label{4.10b}
\end{eqnarray}
can be put in the form
\begin{equation}
\hat{\chi}=\delta \hat{\varphi}+\gamma \hat{\omega}+\partial _{\mu }j^{\mu }.
\label{4.10c}
\end{equation}
Assume that (\ref{4.10c}) holds. Then, by applying $\delta $ on this
equation we infer
\begin{equation}
\delta \hat{\chi}=\gamma \left( -\delta \hat{\omega}\right) +\partial _{\mu
}\left( \delta j^{\mu }\right) .  \label{4.10d}
\end{equation}
On the other hand, if we use the concrete expression (\ref{4.10b}) of $\hat{%
\chi}$, by direct computation we are led to
\begin{eqnarray}
\delta \hat{\chi} &=&\gamma \left( \delta \left( -\left(
k_{D}f_{AB}^{D}-k_{A}k_{B}\right) \bar{\psi}\bar{\psi}^{*}\eta _{\mu
}^{A}\left( \partial ^{\mu }h^{B}-\partial _{\nu }h^{B\mu \nu }\right)
\right) \right.  \nonumber \\
&&+\left( k_{D}f_{AB}^{D}-k_{A}k_{B}\right) \left( \mathrm{i}\bar{\psi}%
\gamma _{\mu }\left( \partial ^{\alpha }\psi \right) \left( \frac{1}{2}%
h^{A\mu \beta }\partial _{[\alpha }^{\left. {}\right. }\eta _{\beta
]}^{B}-\eta ^{A\beta }\partial _{[\alpha }^{\left. {}\right. }h_{\beta
]}^{B\mu }\right) \right.  \nonumber \\
&&\left. \left. -\frac{\mathrm{i}}{8}\bar{\psi}\gamma ^{\mu }\left[ \gamma
^{\alpha },\gamma ^{\beta }\right] \psi \sigma ^{\rho \lambda }\partial
_{[\rho }^{\left. {}\right. }\eta _{\alpha ]}^{A}\partial _{[\lambda
}^{\left. {}\right. }\eta _{\beta ]}^{B}\right) \right)  \nonumber \\
&&+\partial _{\mu }\left( \delta \left( \left(
k_{D}f_{AB}^{D}-k_{A}k_{B}\right) \bar{\psi}\bar{\psi}^{*}\eta ^{A\nu
}\left( \partial ^{\mu }\eta _{\nu }^{B}-\partial _{\nu }\eta ^{B\mu
}\right) \right) \right.  \nonumber \\
&&+\mathrm{i}\left( k_{D}f_{AB}^{D}-k_{A}k_{B}\right) \left( \bar{\psi}%
\gamma ^{\mu }\left( \partial ^{\alpha }\psi \right) \eta ^{A\beta }\partial
_{[\alpha }^{\left. {}\right. }\eta _{\beta ]}^{B}\right.  \nonumber \\
&&\left. \left. +\frac{1}{16}\bar{\psi}\gamma ^{\mu }\left[ \gamma ^{\alpha
},\gamma ^{\beta }\right] \psi \sigma ^{\rho \lambda }\partial _{[\rho
}^{\left. {}\right. }\eta _{\alpha ]}^{A}\partial _{[\lambda }^{\left.
{}\right. }\eta _{\beta ]}^{B}\right) \right) .  \label{4.10e}
\end{eqnarray}
The right-hand side of (\ref{4.10e}) can be written like in the right-hand
side of (\ref{4.10d}) if the following conditions are simultaneously
fulfilled
\begin{eqnarray}
&&\left( k_{D}f_{AB}^{D}-k_{A}k_{B}\right) \left( \mathrm{i}\bar{\psi}\gamma
_{\mu }\left( \partial ^{\alpha }\psi \right) \left( \frac{1}{2}h^{A\mu
\beta }\partial _{[\alpha }^{\left. {}\right. }\eta _{\beta ]}^{B}-\eta
^{A\beta }\partial _{[\alpha }^{\left. {}\right. }h_{\beta ]}^{B\mu }\right)
\right.  \nonumber \\
&&\left. -\frac{\mathrm{i}}{8}\bar{\psi}\gamma ^{\mu }\left[ \gamma ^{\alpha
},\gamma ^{\beta }\right] \psi \sigma ^{\rho \lambda }\partial _{[\rho
}^{\left. {}\right. }\eta _{\alpha ]}^{A}\partial _{[\lambda }^{\left.
{}\right. }\eta _{\beta ]}^{B}\right) =-\delta \hat{\omega}^{\prime },
\label{4.10f}
\end{eqnarray}
\begin{eqnarray}
&&+\mathrm{i}\left( k_{D}f_{AB}^{D}-k_{A}k_{B}\right) \left( \bar{\psi}%
\gamma ^{\mu }\left( \partial ^{\alpha }\psi \right) \eta ^{A\beta }\partial
_{[\alpha }^{\left. {}\right. }\eta _{\beta ]}^{B}\right.  \nonumber \\
&&\left. +\frac{1}{16}\bar{\psi}\gamma ^{\mu }\left[ \gamma ^{\alpha
},\gamma ^{\beta }\right] \psi \sigma ^{\rho \lambda }\partial _{[\rho
}^{\left. {}\right. }\eta _{\alpha ]}^{A}\partial _{[\lambda }^{\left.
{}\right. }\eta _{\beta ]}^{B}\right) =\delta j^{\prime \mu }.  \label{4.10g}
\end{eqnarray}
However, from the action of $\delta $ on the BRST generators we observe that
none of $h^{A\mu \beta }$, $\partial _{[\alpha }^{\left. {}\right. }h_{\beta
]\mu }^{A}$, $\eta _{\beta }^{A}$ and $\partial _{[\lambda }^{\left.
{}\right. }\eta _{\beta ]}^{A}$ are $\delta $-exact. In consequence, the
relations (\ref{4.10f})--(\ref{4.10g}) hold if the equations
\begin{equation}
\bar{\psi}\gamma ^{\mu }\left( \partial _{\alpha }\psi \right) =\delta
\Omega _{\alpha }^{\mu },  \label{5.1x}
\end{equation}
and
\begin{equation}
\bar{\psi}\gamma ^{\mu }\left[ \gamma _{\alpha },\gamma _{\beta }\right]
\psi =\delta \Gamma _{\alpha \beta }^{\mu }  \label{5.2x}
\end{equation}
take place simultaneously. Let us suppose that the relations (\ref{5.1x})--(%
\ref{5.2x}) are indeed satisfied. Acting with $\partial _{\mu }$ on (\ref%
{5.1x})--(\ref{5.2x}) we arrive at
\begin{equation}
\partial _{\mu }\left( \bar{\psi}\gamma ^{\mu }\left( \partial _{\alpha
}\psi \right) \right) =\delta \left( \partial _{\mu }\Omega _{\alpha }^{\mu
}\right) ,  \label{54x}
\end{equation}
and
\begin{equation}
\partial _{\mu }\left( \bar{\psi}\gamma ^{\mu }\left[ \gamma _{\alpha
},\gamma _{\beta }\right] \psi \right) =\delta \left( \partial _{\mu }\Gamma
_{\alpha \beta }^{\mu }\right) .  \label{5.5x}
\end{equation}
On the other hand, by direct computation we arrive at
\begin{eqnarray}
\partial _{\mu }\left( \bar{\psi}\gamma ^{\mu }\left( \partial ^{\alpha
}\psi \right) \right) &=&\delta \left( -\mathrm{i}\left( \psi ^{*}\left(
\partial ^{\alpha }\psi \right) -\bar{\psi}\left( \partial ^{\alpha }\bar{%
\psi}^{*}\right) \right) \right) ,  \label{4.11a} \\
\partial _{\mu }\left( \bar{\psi}\gamma ^{\mu }\left[ \gamma _{\alpha
},\gamma _{\beta }\right] \psi \right) &=&\delta \left( \mathrm{i}M_{\alpha
\beta }\right) -4\bar{\psi}\gamma _{\left[ \alpha \right. }\partial _{\left.
\beta \right] }\psi .  \label{4.11b}
\end{eqnarray}
The right-hand sides of (\ref{4.11a})--(\ref{4.11b}) are not of the same
type like the corresponding ones in (\ref{54x})--(\ref{5.5x}). This means
that the relations (\ref{5.1x})--(\ref{5.2x}) are not valid, and therefore
neither are (\ref{4.10f})--(\ref{4.10g}). As a consequence, $\hat{\chi}$
must vanish, which further implies that
\begin{equation}
k_{D}f_{AB}^{D}-k_{A}k_{B}=0.  \label{4.12c}
\end{equation}
Using (\ref{4.12c}) and (\ref{4.21}) we obtain that for $A\neq B$
\begin{equation}
k_{A}k_{B}=0,  \label{zwx}
\end{equation}
which shows that the Dirac fields can couple to only one graviton, which
proves the assertion from the beginning of this section.

\section{Conclusion}

To conclude with, in this paper we have investigated the couplings between a
collection of massless spin-two fields (described in the free limit by a sum
of Pauli-Fierz actions) and a Dirac field using the powerful setting based
on local BRST cohomology. Initially, we have shown that, if we decompose the
metric like $g_{\mu \nu }=\sigma _{\mu \nu }+gh_{\mu \nu }$, then we can
couple Dirac spinors to $h_{\mu \nu }$ in the space of formal series with
the maximum derivative order equal to one in $h_{\mu \nu }$, such that the
final results agree with the usual couplings between the spin-1/2 and the
massless spin-two field in the vierbein formulation. Based on this result,
we have proved, under the hypotheses of locality, smoothness of the
interactions in the coupling constant, Poincar\'{e} invariance, (background)
Lorentz invariance and the preservation of the number of derivatives on each
field, that there are no consistent cross-interactions among different
gravitons in the presence of a Dirac field.

\section*{Acknowledgment}

Three of the authors (C.B., E.M.C. and S.O.S.) are partially supported by
the European Commission FP6 program MRTN-CT-2004-005104 and by the type A
grant 305/2004 with the Romanian National Council for Academic Scientific
Research (C.N.C.S.I.S.) and the Romanian Ministry of Education and Research
(M.E.C.). One of the authors (A.C.L.) is supported by the World Federation
of Scientists (WFS) National Scholarship Programme. Useful discussions with
Glenn Barnich and Ion I. Cotaescu are also acknowledged.

\appendix

\section{Proof of a statement made in subsection 4.2\label{appa}}

Here, we prove that a term of the type
\begin{equation}
\tilde{a}_{1}^{\left( \mathrm{int}\right) }=h^{*\mu \nu }\eta _{\mu }F_{\nu
}\left( \bar{\psi},\psi \right)  \label{PFD3.24x}
\end{equation}
is consistent in antighost number zero,
\begin{equation}
\delta \tilde{a}_{1}^{\left( \mathrm{int}\right) }+\gamma \tilde{a}%
_{0}^{\left( \mathrm{int}\right) }=\partial _{\mu }\rho ^{\mu },
\label{PFD3.24y}
\end{equation}
if and only if
\begin{equation}
F_{\nu }\left( \bar{\psi},\psi \right) =\partial _{\nu }F\left( \bar{\psi}%
,\psi \right) .  \label{PFD3.24z}
\end{equation}
Indeed, by applying $\delta $ on $\tilde{a}_{1}^{\left( \mathrm{int}\right)
} $ we obtain that
\begin{equation}
\delta \tilde{a}_{1}^{\left( \mathrm{int}\right) }=-2H^{\mu \nu }\eta _{\mu
}F_{\nu }\left( \bar{\psi},\psi \right) .  \label{PFD3.24q}
\end{equation}
It is easy to see that, if $F_{\nu }\left( \bar{\psi},\psi \right) $ if the
form (\ref{PFD3.24z}), then (\ref{PFD3.24q}) implies
\begin{equation}
\delta \tilde{a}_{1}^{\left( \mathrm{int}\right) }=\gamma \left( H^{\mu \nu
}h_{\mu \nu }F\left( \bar{\psi},\psi \right) \right) +\partial _{\mu }\left(
-2H^{\mu \nu }\eta _{\nu }F\left( \bar{\psi},\psi \right) \right) ,
\label{PFD3.24w}
\end{equation}
and therefore $\tilde{a}_{1}^{\left( \mathrm{int}\right) }$ indeed checks an
equation of the type (\ref{PFD3.24y}). Let us suppose now that $\tilde{a}%
_{1}^{\left( \mathrm{int}\right) }$ satisfies the equation (\ref{PFD3.24y}).
Inserting the relations
\begin{equation}
\gamma \tilde{a}_{0}^{\left( \mathrm{int}\right) }=2\frac{\delta \tilde{a}%
_{0}^{\left( \mathrm{int}\right) }}{\delta h_{\mu \nu }}\partial _{\mu }\eta
_{\nu }+\partial _{\mu }t^{\mu },  \label{PFD3.24p}
\end{equation}
and (\ref{PFD3.24q}) in (\ref{PFD3.24y}), we get that
\begin{equation}
-2H^{\mu \nu }\eta _{\mu }F_{\nu }\left( \bar{\psi},\psi \right) +2\frac{%
\delta \tilde{a}_{0}^{\left( \mathrm{int}\right) }}{\delta h_{\mu \nu }}%
\partial _{\nu }\eta _{\mu }=\partial _{\mu }p^{\mu }.  \label{PFD3.24l}
\end{equation}
The left-hand side of the last relation reduces to a total derivative if
\begin{equation}
H^{\mu \nu }F_{\nu }\left( \bar{\psi},\psi \right) =-\partial _{\nu }\frac{%
\delta \tilde{a}_{0}^{\left( \mathrm{int}\right) }}{\delta h_{\mu \nu }}.
\label{PFD3.24k}
\end{equation}
In order to investigate under what conditions the left-hand side of (\ref%
{PFD3.24k}) also provides a total derivative, we start from the fact that
\begin{equation}
H^{\mu \nu }=\partial _{\alpha }\partial _{\beta }\phi ^{\mu \alpha \nu
\beta },  \label{PFD3.24m}
\end{equation}
where
\begin{eqnarray}
\phi ^{\mu \alpha \nu \beta } &=&\frac{1}{2}\left( -h^{\mu \nu }\sigma
^{\alpha \beta }+h^{\alpha \nu }\sigma ^{\mu \beta }+h^{\mu \beta }\sigma
^{\alpha \nu }-h^{\alpha \beta }\sigma ^{\mu \nu }\right.  \nonumber \\
&&\left. +h\left( \sigma ^{\mu \nu }\sigma ^{\alpha \beta }-\sigma ^{\mu
\beta }\sigma ^{\alpha \nu }\right) \right) .  \label{PFD3.24n}
\end{eqnarray}
By means of (\ref{PFD3.24m}) we further deduce that
\begin{eqnarray}
H^{\mu \nu }F_{\nu } &=&\partial _{\nu }\left( \partial _{\beta }\phi ^{\mu
\nu \alpha \beta }F_{\alpha }-\phi ^{\mu \beta \alpha \nu }\partial _{\beta
}F_{\alpha }\right)  \nonumber \\
&&+\frac{1}{2}\phi ^{\mu \alpha \nu \beta }\partial _{\alpha }\partial _{%
\left[ \beta \right. }F_{\left. \nu \right] }.  \label{PFD3.24v}
\end{eqnarray}
Thus, the right-hand side of (\ref{PFD3.24v}) gives a total derivative if
and only if
\[
\phi ^{\mu \alpha \nu \beta }\partial _{\alpha }\partial _{\left[ \beta
\right. }F_{\left. \nu \right] }=0,
\]
which further yields $F_{\nu }=\partial _{\nu }F$. This completes the proof.

\section{Complete computation of the second-order deformation\label{appb}}

In this appendix we are interested in determining the complete expression of
the second-order deformation for the master equation, which is known to be
subject to the equation (\ref{PFD2.6}). Proceeding in the same manner like
during the first-order deformation procedure, we can write the second-order
deformation of the master equation like the sum between the Pauli-Fierz and
the interacting parts
\begin{equation}
S_{2}=S_{2}^{\left( \mathrm{PF}\right) }+S_{2}^{\left( \mathrm{int}\right) }.
\label{PFD4.2}
\end{equation}
The piece $S_{2}^{\left( \mathrm{PF}\right) }$ describes the second-order
deformation in the Pauli-Fierz sector and we will not insist on it since we
are merely interested in the cross-couplings. The term $S_{2}^{\left(
\mathrm{int}\right) }$ results as solution to the equation
\begin{equation}
\frac{1}{2}\left( S_{1},S_{1}\right) ^{\left( \mathrm{int}\right)
}+sS_{2}^{\left( \mathrm{int}\right) }=0,  \label{PFD4.3}
\end{equation}
where
\begin{equation}
\left( S_{1},S_{1}\right) ^{\left( \mathrm{int}\right) }=\left(
S_{1}^{\left( \mathrm{int}\right) },S_{1}^{\left( \mathrm{int}\right)
}\right) +2\left( S_{1}^{\left( \mathrm{PF}\right) },S_{1}^{\left( \mathrm{%
int}\right) }\right) .  \label{PFD4.4}
\end{equation}
If we denote by $\Delta ^{\left( \mathrm{int}\right) }$ and $b^{\left(
\mathrm{int}\right) }$ the non-integrated densities of $\left(
S_{1},S_{1}\right) ^{\left( \mathrm{int}\right) }$ and respectively of $%
S_{2}^{\left( \mathrm{int}\right) }$, the local form of (\ref{PFD4.3})
becomes
\begin{equation}
\Delta ^{\left( \mathrm{int}\right) }=-2sb^{\left( \mathrm{int}\right)
}+\partial _{\mu }n^{\mu },  \label{PFD4.5}
\end{equation}
with
\begin{equation}
\mathrm{gh}\left( \Delta ^{\left( \mathrm{int}\right) }\right) =1,\quad
\mathrm{gh}\left( b^{\left( \mathrm{int}\right) }\right) =0,\quad \mathrm{gh}%
\left( n^{\mu }\right) =1,  \label{PFD45a}
\end{equation}
for some local currents $n^{\mu }$. Direct computation shows that $\Delta
^{\left( \mathrm{int}\right) }$ decomposes like
\begin{equation}
\Delta ^{\left( \mathrm{int}\right) }=\Delta _{0}^{\left( \mathrm{int}%
\right) }+\Delta _{1}^{\left( \mathrm{int}\right) },\quad \mathrm{agh}\left(
\Delta _{I}^{\left( \mathrm{int}\right) }\right) =I,\quad I=0,1,
\label{PFDx}
\end{equation}
with
\begin{eqnarray}
\Delta _{1}^{\left( \mathrm{int}\right) } &=&\gamma \left( k^{2}\left( \psi
^{*}\left( \partial ^{\alpha }\psi \right) +\left( \partial ^{\alpha }\bar{%
\psi}\right) \bar{\psi}^{*}\right) \eta ^{\beta }h_{\alpha \beta }\right.
\nonumber \\
&&\left. +\frac{k}{16}M^{\alpha \beta }\left( \left( 1-2k\right) \eta
^{\sigma }\partial _{[\alpha }h_{\beta ]\sigma }+\frac{1}{2}h_{\sigma
[\alpha }\left( \partial _{\beta ]}\eta ^{\sigma }-\partial ^{\sigma }\eta
_{\beta ]}\right) \right) \right)  \nonumber \\
&&+k\left( 1-k\right) \left( \psi ^{*}\left( \partial _{\alpha }\psi \right)
+\left( \partial _{\alpha }\bar{\psi}\right) \bar{\psi}^{*}\right) \eta
_{\beta }\partial ^{\left[ \alpha \right. }\eta ^{\beta ]}  \nonumber \\
&&-\frac{k}{16}\left( 1-k\right) M^{\alpha \beta }\sigma ^{\mu \nu }\partial
_{[\alpha }\eta _{\mu ]}\partial _{[\beta }\eta _{\nu ]},  \label{PFD4.12a}
\end{eqnarray}
and
\begin{eqnarray}
\Delta _{0}^{\left( \mathrm{int}\right) } &=&k\mathcal{L}_{0}^{(\mathrm{D}%
)}\left( h_{\alpha \beta }\partial ^{\alpha }\eta ^{\beta }+\eta ^{\beta
}\left( \partial _{\beta }h-\partial ^{\alpha }h_{\alpha \beta }\right)
\right)  \nonumber \\
&&+\frac{\mathrm{i}k}{2}\bar{\psi}\gamma ^{(\alpha }\left( \partial ^{\beta
)}\psi \right) \left( -h_{\sigma \alpha }\partial _{\beta }\eta ^{\sigma
}+\eta ^{\sigma }\left( \partial _{(\alpha }h_{\beta )\sigma }-2\partial
_{\sigma }h_{\alpha \beta }\right) \right)  \nonumber \\
&&+\frac{\mathrm{i}k}{8}\bar{\psi}\gamma ^{\mu }[\gamma ^{\alpha },\gamma
^{\beta }]\psi \partial _{\alpha }\left( \eta ^{\sigma }\left( \partial
_{(\beta }h_{\mu )\sigma }-2\partial _{\sigma }h_{\mu \beta }\right) -\left(
\partial _{(\beta }\eta ^{\sigma }\right) h_{\mu )\sigma }\right)  \nonumber
\\
&&+k^{2}h\left( \left( \partial ^{\sigma }\mathcal{L}_{0}^{(\mathrm{D}%
)}\right) \eta _{\sigma }+\bar{\psi}\mathrm{i}\gamma ^{\alpha }\left(
\partial ^{\beta }\psi \right) \partial _{\alpha }\eta _{\beta }\right)
\nonumber \\
&&-\mathrm{i}k^{2}h_{\alpha \beta }\left( \left( \partial ^{\sigma }\bar{\psi%
}\right) \gamma ^{\alpha }\left( \partial ^{\beta }\psi \right) \eta
_{\sigma }+\bar{\psi}\gamma ^{\alpha }\partial ^{\beta }\left( \partial
^{\sigma }\psi \eta _{\sigma }\right) \right)  \nonumber \\
&&-\frac{\mathrm{i}k^{2}}{8}\left( \partial ^{\sigma }\left( \bar{\psi}%
\gamma ^{\mu }[\gamma ^{\alpha },\gamma ^{\beta }]\psi \right) \right) \eta
_{\sigma }\partial _{[\alpha }h_{\beta ]\mu }+\frac{\mathrm{i}k^{2}}{16}\bar{%
\psi}\gamma ^{\mu }[\gamma ^{\alpha },\gamma ^{\beta }]\psi h\partial _{\mu
}\left( \partial _{[\alpha }\eta _{\beta ]}\right)  \nonumber \\
&&+\frac{\mathrm{i}k^{2}}{16}\left( \bar{\psi}[\gamma ^{\mu },\gamma ^{\nu
}]\gamma ^{\alpha }\left( \partial ^{\beta }\psi \right) h_{\alpha \beta
}\partial _{[\mu }\eta _{\nu ]}-\bar{\psi}\gamma ^{\alpha }[\gamma ^{\mu
},\gamma ^{\nu }]\left( \partial ^{\beta }\left( \psi \partial _{[\mu }\eta
_{\nu ]}\right) \right) h_{\alpha \beta }\right)  \nonumber \\
&&+\frac{\mathrm{i}k^{2}}{128}\bar{\psi}\left( [\gamma ^{\rho },\gamma
^{\lambda }]\gamma ^{\mu }[\gamma ^{\alpha },\gamma ^{\beta }]-\gamma ^{\mu
}[\gamma ^{\alpha },\gamma ^{\beta }][\gamma ^{\rho },\gamma ^{\lambda
}]\right) \psi \left( \partial _{[\alpha }h_{\beta ]\mu }\right) \partial
_{[\rho }\eta _{\lambda ]}  \nonumber \\
&&+\gamma \left( -k\left( f\left( \bar{\psi},\psi \right) +\left( \partial
_{\mu }\bar{\psi}\right) g_{1}^{\mu }\left( \bar{\psi},\psi \right)
+g_{2}^{\mu }\left( \bar{\psi},\psi \right) \left( \partial _{\mu }\psi
\right) \right) h\right.  \nonumber \\
&&\left. +kh_{\mu \nu }\left( \left( \partial ^{\nu }\bar{\psi}\right)
g_{1}^{\mu }\left( \bar{\psi},\psi \right) +g_{2}^{\mu }\left( \bar{\psi}%
,\psi \right) \left( \partial ^{\nu }\psi \right) \right) \right)  \nonumber
\\
&&+\frac{k}{8}\left( \frac{\partial ^{R}f}{\partial \psi }\left[ \gamma
^{\mu },\gamma ^{\nu }\right] \psi -\bar{\psi}\left[ \gamma ^{\mu },\gamma
^{\nu }\right] \frac{\partial ^{L}f}{\partial \bar{\psi}}\right) \partial
_{[\mu }\eta _{\nu ]}  \nonumber \\
&&+\frac{k}{8}\partial _{[\mu }\eta _{\nu ]}\left( \bar{\psi}\left[ \gamma
^{\mu },\gamma ^{\nu }\right] \left( \partial _{\rho }g_{1}^{\rho }\right)
-\left( \partial _{\rho }g_{2}^{\rho }\right) \left[ \gamma ^{\mu },\gamma
^{\nu }\right] \psi \right.  \nonumber \\
&&+\frac{\partial ^{R}\left( \left( \partial _{\rho }\bar{\psi}\right)
g_{1}^{\rho }+g_{2}^{\rho }\left( \partial _{\rho }\psi \right) \right) }{%
\partial \psi }\left[ \gamma ^{\mu },\gamma ^{\nu }\right] \psi  \nonumber \\
&&-\bar{\psi}\left[ \gamma ^{\mu },\gamma ^{\nu }\right] \frac{\partial
^{L}\left( \left( \partial _{\rho }\bar{\psi}\right) g_{1}^{\rho
}+g_{2}^{\rho }\left( \partial _{\rho }\psi \right) \right) }{\partial \bar{%
\psi}}  \nonumber \\
&&\left. -4\left( \left( \partial _{\left. {}\right. }^{\left[ \mu \right. }%
\bar{\psi}\right) g_{1}^{\left. \nu \right] }-g_{2}^{\left[ \mu \right.
}\left( \partial _{\left. {}\right. }^{\left. \nu \right] }\psi \right)
\right) \right) ,  \label{xxx}
\end{eqnarray}
where we used the notation
\begin{equation}
M^{\alpha \beta }=\bar{\psi}\left[ \gamma ^{\alpha },\gamma ^{\beta }\right]
\bar{\psi}^{*}-\psi ^{*}\left[ \gamma ^{\alpha },\gamma ^{\beta }\right]
\psi .  \label{PFD4.12b}
\end{equation}

Since the first-order deformation in the interacting sector starts in
antighost number one, we can take, without loss of generality, the
corresponding second-order deformation to start in antighost number two
\begin{eqnarray}
b^{\left( \mathrm{int}\right) } &=&b_{0}^{\left( \mathrm{int}\right)
}+b_{1}^{\left( \mathrm{int}\right) }+b_{2}^{\left( \mathrm{int}\right)
},\quad \mathrm{agh}\left( b_{I}^{\left( \mathrm{int}\right) }\right)
=I,\quad I=0,1,2,  \label{PFD4.7} \\
n^{\mu } &=&n_{0}^{\mu }+n_{1}^{\mu }+n_{2}^{\mu },\quad \mathrm{agh}\left(
n_{I}^{\mu }\right) =I,\quad I=0,1,2.  \label{PFD4.8}
\end{eqnarray}
By projecting the equation (\ref{PFD4.5}) on various antighost numbers, we
obtain
\begin{eqnarray}
\gamma b_{2}^{\left( \mathrm{int}\right) } &=&\partial _{\mu }\left( \frac{1%
}{2}n_{2}^{\mu }\right) ,  \label{PFD4.9x} \\
\Delta _{1}^{\left( \mathrm{int}\right) } &=&-2\left( \delta b_{2}^{\left(
\mathrm{int}\right) }+\gamma b_{1}^{\left( \mathrm{int}\right) }\right)
+\partial _{\mu }n_{1}^{\mu },  \label{PFD4.9} \\
\Delta _{0}^{\left( \mathrm{int}\right) } &=&-2\left( \delta b_{1}^{\left(
\mathrm{int}\right) }+\gamma b_{0}^{\left( \mathrm{int}\right) }\right)
+\partial _{\mu }n_{0}^{\mu }.  \label{PFD4.10}
\end{eqnarray}
The equation (\ref{PFD4.9x}) can always be replaced, by adding trivial
terms, with
\begin{equation}
\gamma b_{2}^{\left( \mathrm{int}\right) }=0.  \label{PFD4.10x}
\end{equation}
Looking at $\Delta _{1}^{\left( \mathrm{int}\right) }$ given in (\ref%
{PFD4.12a}), it results that it can be written like in (\ref{PFD4.9}) if
\begin{eqnarray}
\chi &=&k\left( 1-k\right) \left( \left( \psi ^{*}\left( \partial _{\alpha
}\psi \right) +\left( \partial _{\alpha }\bar{\psi}\right) \bar{\psi}%
^{*}\right) \eta _{\beta }\partial ^{\left[ \alpha \right. }\eta ^{\beta
]}\right.  \nonumber \\
&&\left. -\frac{1}{16}M^{\alpha \beta }\sigma ^{\mu \nu }\partial _{[\alpha
}\eta _{\mu ]}\partial _{[\beta }\eta _{\nu ]}\right) ,  \label{PFD4.10y}
\end{eqnarray}
can be expressed like
\begin{equation}
\chi =\delta \varphi +\gamma \omega +\partial _{\mu }l^{\mu }.
\label{PFD4.10z}
\end{equation}
Supposing that (\ref{PFD4.10z}) holds and applying $\delta $ on it, we infer
that
\begin{equation}
\delta \chi =\gamma \left( -\delta \omega \right) +\partial _{\mu }\left(
\delta l^{\mu }\right) .  \label{PFD4.10w}
\end{equation}
On the other hand, using the concrete expression of $\chi $, we have that
\begin{eqnarray}
\delta \chi &=&\gamma \left( \delta \left( k\left( 1-k\right) \bar{\psi}\bar{%
\psi}^{*}\left( \partial _{\mu }h^{\mu \beta }-\partial _{\beta }h\right)
\eta _{\beta }\right) \right)  \nonumber \\
&&+\partial _{\mu }\left( \delta \left( k\left( 1-k\right) \bar{\psi}\bar{%
\psi}^{*}\eta _{\beta }\partial ^{\left[ \mu \right. }\eta ^{\beta ]}\right)
\right)  \nonumber \\
&&+\gamma \left( \mathrm{i}k\left( 1-k\right) \bar{\psi}\gamma ^{\mu }\left(
\partial _{\alpha }\psi \right) \left( \frac{1}{2}h_{\mu \beta }\partial ^{%
\left[ \alpha \right. }\eta ^{\beta ]}-\left( \partial ^{\left[ \alpha
\right. }h^{\beta ]\lambda }\right) \sigma _{\lambda \mu }\eta _{\beta
}\right) \right.  \nonumber \\
&&\left. +\frac{\mathrm{i}}{8}k\left( 1-k\right) \bar{\psi}\gamma ^{\mu }%
\left[ \gamma ^{\alpha },\gamma ^{\beta }\right] \psi \partial _{[\alpha
}h_{\lambda ]\mu }\partial _{[\beta }\eta _{\nu ]}\sigma ^{\lambda \nu
}\right)  \nonumber \\
&&+\partial _{\mu }\left( \mathrm{i}k\left( 1-k\right) \bar{\psi}\gamma
^{\mu }\left( \partial _{\alpha }\psi \right) \eta _{\beta }\partial ^{\left[
\alpha \right. }\eta ^{\beta ]}\right.  \nonumber \\
&&\left. +\frac{\mathrm{i}}{16}k\left( 1-k\right) \bar{\psi}\gamma ^{\mu }%
\left[ \gamma ^{\alpha },\gamma ^{\beta }\right] \psi \partial _{[\alpha
}\eta _{\lambda ]}\partial _{[\beta }\eta _{\nu ]}\sigma ^{\lambda \nu
}\right) .  \label{PFD4.10q}
\end{eqnarray}
The right-hand side of (\ref{PFD4.10q}) can be written like in the
right-hand side of (\ref{PFD4.10w}) if the following conditions are
simultaneously satisfied
\begin{eqnarray}
-\delta \omega ^{\prime } &=&\bar{\psi}\gamma ^{\mu }\left( \partial
_{\alpha }\psi \right) \left( \frac{1}{2}h_{\mu \beta }\partial ^{\left[
\alpha \right. }\eta ^{\beta ]}-\left( \partial ^{\left[ \alpha \right.
}h^{\beta ]\lambda }\right) \sigma _{\lambda \mu }\eta _{\beta }\right)
\nonumber \\
&&+\frac{1}{8}\bar{\psi}\gamma ^{\mu }\left[ \gamma ^{\alpha },\gamma
^{\beta }\right] \psi \partial _{[\alpha }h_{\lambda ]\mu }\partial _{[\beta
}\eta _{\nu ]}\sigma ^{\lambda \nu },  \label{PFD4.10v}
\end{eqnarray}
\begin{eqnarray}
\delta l^{\prime \mu } &=&\bar{\psi}\gamma ^{\mu }\left( \partial _{\alpha
}\psi \right) \eta _{\beta }\partial ^{\left[ \alpha \right. }\eta ^{\beta ]}
\nonumber \\
&&+\frac{1}{16}\bar{\psi}\gamma ^{\mu }\left[ \gamma ^{\alpha },\gamma
^{\beta }\right] \psi \partial _{[\alpha }\eta _{\lambda ]}\partial _{[\beta
}\eta _{\nu ]}\sigma ^{\lambda \nu }.  \label{PFD4.10u}
\end{eqnarray}
Since none of the quantities $h_{\mu \beta }$, $\partial ^{\left[ \alpha
\right. }h^{\beta ]\lambda }$, $\eta _{\beta }$ or $\partial ^{\left[ \alpha
\right. }\eta ^{\beta ]}$ are $\delta $-exact, the last relations hold if
the equations
\begin{equation}
\bar{\psi}\gamma ^{\mu }\left( \partial _{\alpha }\psi \right) =\delta
\Omega _{\;\;\alpha }^{\mu },\;\bar{\psi}\gamma ^{\mu }\left[ \gamma
_{\alpha },\gamma _{\beta }\right] \psi =\delta \Gamma _{\;\;\alpha \beta
}^{\mu }  \label{PFD4.10p}
\end{equation}
take place simultaneously. Assuming that both the equations (\ref{PFD4.10p})
are valid, they further give
\begin{eqnarray}
\partial _{\mu }\left( \bar{\psi}\gamma ^{\mu }\left( \partial _{\alpha
}\psi \right) \right) &=&\delta \left( \partial _{\mu }\Omega _{\;\;\alpha
}^{\mu }\right) ,  \label{PFD4.10xx} \\
\partial _{\mu }\left( \bar{\psi}\gamma ^{\mu }\left[ \gamma _{\alpha
},\gamma _{\beta }\right] \psi \right) &=&\delta \left( \partial _{\mu
}\Gamma _{\;\;\alpha \beta }^{\mu }\right) .  \label{PFD4.10zx}
\end{eqnarray}
On the other hand, by direct computation we obtain that
\begin{eqnarray}
\partial _{\mu }\left( \bar{\psi}\gamma ^{\mu }\left( \partial _{\alpha
}\psi \right) \right) &=&\delta \left( -\mathrm{i}\left( \psi ^{*}\left(
\partial _{\alpha }\psi \right) -\bar{\psi}\left( \partial _{\alpha }\bar{%
\psi}^{*}\right) \right) \right) ,  \label{PFD4.10l} \\
\partial _{\mu }\left( \bar{\psi}\gamma ^{\mu }\left[ \gamma _{\alpha
},\gamma _{\beta }\right] \psi \right) &=&\delta \left( \mathrm{i}M_{\alpha
\beta }\right) -4\bar{\psi}\gamma _{\left[ \alpha \right. }\left( \partial
_{\left. \beta \right] }\psi \right) ,  \label{PFD4.10ww}
\end{eqnarray}
so the right-hand sides of (\ref{PFD4.10l})--(\ref{PFD4.10ww}) cannot be
written like in the right-hand sides of (\ref{PFD4.10xx})--(\ref{PFD4.10zx}%
). This means that the relations (\ref{PFD4.10p}) are not valid, and
therefore neither are (\ref{PFD4.10v})--(\ref{PFD4.10u}). As a consequence, $%
\chi $ must vanish, and hence we must set
\begin{equation}
k\left( 1-k\right) =0.  \label{PFD4.12c}
\end{equation}
Using (\ref{PFD4.12c}), we conclude that
\begin{equation}
k=1.  \label{PFD4.14}
\end{equation}
Inserting (\ref{PFD4.14}) in (\ref{PFD4.12a}), we obtain that
\begin{eqnarray}
\Delta _{1}^{\left( \mathrm{int}\right) } &=&\gamma \left( \left( \psi
^{*}\left( \partial ^{\alpha }\psi \right) +\left( \partial ^{\alpha }\bar{%
\psi}\right) \bar{\psi}^{*}\right) \eta ^{\beta }h_{\alpha \beta }\right.
\nonumber \\
&&\left. -\frac{1}{16}M^{\alpha \beta }\left( \eta ^{\sigma }\partial
_{[\alpha }h_{\beta ]\sigma }-\frac{1}{2}h_{\sigma [\alpha }\left( \partial
_{\beta ]}\eta ^{\sigma }-\partial ^{\sigma }\eta _{\beta ]}\right) \right)
\right) .  \label{PFD4.15}
\end{eqnarray}

Comparing (\ref{PFD4.15}) with (\ref{PFD4.9}), we find that
\begin{equation}
b_{2}^{\left( \mathrm{int}\right) }=0,  \label{PFD4.15x}
\end{equation}
\begin{eqnarray}
b_{1}^{\left( \mathrm{int}\right) } &=&-\frac{1}{2}\left( \psi ^{*}\left(
\partial ^{\alpha }\psi \right) +\left( \partial ^{\alpha }\bar{\psi}\right)
\bar{\psi}^{*}\right) \eta ^{\beta }h_{\alpha \beta }+\frac{1}{32}\left(
\bar{\psi}\left[ \gamma ^{\alpha },\gamma ^{\beta }\right] \bar{\psi}%
^{*}\right.  \nonumber \\
&&\left. -\psi ^{*}\left[ \gamma ^{\alpha },\gamma ^{\beta }\right] \psi
\right) \left( \eta ^{\sigma }\partial _{[\alpha }h_{\beta ]\sigma }-\frac{1%
}{2}h_{\sigma [\alpha }\left( \partial _{\beta ]}\eta ^{\sigma }-\partial
^{\sigma }\eta _{\beta ]}\right) \right) .  \label{PFD4.16}
\end{eqnarray}
Substituting (\ref{PFD4.14}) in (\ref{xxx}) and using (\ref{PFD4.16}), we
deduce
\begin{eqnarray}
\Delta _{0}^{\left( \mathrm{int}\right) }+2\delta b_{1}^{\left( \mathrm{int}%
\right) } &=&\partial _{\mu }n_{0}^{\mu }+\gamma \left( \frac{\mathrm{i}}{2}%
\bar{\psi}\gamma ^{\mu }\partial ^{\nu }\psi \left( hh_{\mu \nu }-\frac{3}{2}%
h_{\mu \sigma }h_{\nu }^{\sigma }\right) \right.  \nonumber \\
&&+\frac{1}{2}\left( \bar{\psi}\mathrm{i}\gamma ^{\mu }\left( \partial _{\mu
}\psi \right) -m\bar{\psi}\psi \right) \left( h_{\alpha \beta }h^{\alpha
\beta }-\frac{1}{2}h^{2}\right)  \nonumber \\
&&+\frac{\mathrm{i}}{16}\bar{\psi}\gamma ^{\mu }\left[ \gamma ^{\alpha
},\gamma ^{\beta }\right] \psi \left( h\partial _{[\alpha }h_{\beta ]\mu
}-h_{\mu }^{\sigma }\partial _{[\alpha }h_{\beta ]\sigma }\right.  \nonumber
\\
&&\left. +h_{\alpha }^{\sigma }\left( 2\partial _{[\beta }h_{\sigma ]\mu
}+\partial _{\mu }h_{\beta \sigma }\right) \right)  \nonumber \\
&&-\left( f\left( \bar{\psi},\psi \right) +\left( \partial _{\mu }\bar{\psi}%
\right) g_{1}^{\mu }\left( \bar{\psi},\psi \right) +g_{2}^{\mu }\left( \bar{%
\psi},\psi \right) \left( \partial _{\mu }\psi \right) \right) h  \nonumber
\\
&&\left. +h_{\mu \nu }\left( \left( \partial ^{\nu }\bar{\psi}\right)
g_{1}^{\mu }\left( \bar{\psi},\psi \right) +g_{2}^{\mu }\left( \bar{\psi}%
,\psi \right) \left( \partial ^{\nu }\psi \right) \right) \right)  \nonumber
\\
&&+\frac{1}{8}\left( \frac{\partial ^{R}f}{\partial \psi }\left[ \gamma
^{\mu },\gamma ^{\nu }\right] \psi -\bar{\psi}\left[ \gamma ^{\mu },\gamma
^{\nu }\right] \frac{\partial ^{L}f}{\partial \bar{\psi}}\right) \partial
_{[\mu }\eta _{\nu ]}  \nonumber \\
&&+\frac{1}{8}\partial _{[\mu }\eta _{\nu ]}\left( \bar{\psi}\left[ \gamma
^{\mu },\gamma ^{\nu }\right] \left( \partial _{\rho }g_{1}^{\rho }\right)
-\left( \partial _{\rho }g_{2}^{\rho }\right) \left[ \gamma ^{\mu },\gamma
^{\nu }\right] \psi \right.  \nonumber \\
&&+\frac{\partial ^{R}\left( \left( \partial _{\rho }\bar{\psi}\right)
g_{1}^{\rho }+g_{2}^{\rho }\left( \partial _{\rho }\psi \right) \right) }{%
\partial \psi }\left[ \gamma ^{\mu },\gamma ^{\nu }\right] \psi  \nonumber \\
&&-\bar{\psi}\left[ \gamma ^{\mu },\gamma ^{\nu }\right] \frac{\partial
^{L}\left( \left( \partial _{\rho }\bar{\psi}\right) g_{1}^{\rho
}+g_{2}^{\rho }\left( \partial _{\rho }\psi \right) \right) }{\partial \bar{%
\psi}}  \nonumber \\
&&\left. -4\left( \left( \partial _{\left. {}\right. }^{\left[ \mu \right. }%
\bar{\psi}\right) g_{1}^{\left. \nu \right] }-g_{2}^{\left[ \mu \right.
}\left( \partial _{\left. {}\right. }^{\left. \nu \right] }\psi \right)
\right) \right) .  \label{PFD4.17}
\end{eqnarray}
The right-hand side of (\ref{PFD4.17}) can be written like in (\ref{PFD4.10}%
) if
\begin{eqnarray}
&&\frac{1}{8}\left( \frac{\partial ^{R}f}{\partial \psi }\left[ \gamma ^{\mu
},\gamma ^{\nu }\right] \psi -\bar{\psi}\left[ \gamma ^{\mu },\gamma ^{\nu }%
\right] \frac{\partial ^{L}f}{\partial \bar{\psi}}\right) \partial _{[\mu
}\eta _{\nu ]}  \nonumber \\
&&+\frac{1}{8}\partial _{[\mu }\eta _{\nu ]}\left( \bar{\psi}\left[ \gamma
^{\mu },\gamma ^{\nu }\right] \left( \partial _{\rho }g_{1}^{\rho }\right)
-\left( \partial _{\rho }g_{2}^{\rho }\right) \left[ \gamma ^{\mu },\gamma
^{\nu }\right] \psi \right.  \nonumber \\
&&+\frac{\partial ^{R}\left( \left( \partial _{\rho }\bar{\psi}\right)
g_{1}^{\rho }+g_{2}^{\rho }\left( \partial _{\rho }\psi \right) \right) }{%
\partial \psi }\left[ \gamma ^{\mu },\gamma ^{\nu }\right] \psi  \nonumber \\
&&-\bar{\psi}\left[ \gamma ^{\mu },\gamma ^{\nu }\right] \frac{\partial
^{L}\left( \left( \partial _{\rho }\bar{\psi}\right) g_{1}^{\rho
}+g_{2}^{\rho }\left( \partial _{\rho }\psi \right) \right) }{\partial \bar{%
\psi}}  \nonumber \\
&&\left. -4\left( \left( \partial _{\left. {}\right. }^{\left[ \mu \right. }%
\bar{\psi}\right) g_{1}^{\left. \nu \right] }-g_{2}^{\left[ \mu \right.
}\left( \partial _{\left. {}\right. }^{\left. \nu \right] }\psi \right)
\right) \right) =\gamma \theta +\partial _{\mu }\rho ^{\mu }.
\label{PFD4.18}
\end{eqnarray}
The term $\frac{1}{8}\left( \frac{\partial ^{R}f}{\partial \psi }\left[
\gamma ^{\mu },\gamma ^{\nu }\right] \psi -\bar{\psi}\left[ \gamma ^{\mu
},\gamma ^{\nu }\right] \frac{\partial ^{L}f}{\partial \bar{\psi}}\right)
\partial _{[\mu }\eta _{\nu ]}$ is neither $\gamma $-exact nor a total
derivative (as $f\left( \bar{\psi},\psi \right) $ has no derivatives), and
hence we must require that
\begin{equation}
\frac{\partial ^{R}f}{\partial \psi }\left[ \gamma ^{\mu },\gamma ^{\nu }%
\right] \psi -\bar{\psi}\left[ \gamma ^{\mu },\gamma ^{\nu }\right] \frac{%
\partial ^{L}f}{\partial \bar{\psi}}=0.  \label{PFD4.18x}
\end{equation}
The solution to (\ref{PFD4.18x}) reads as
\begin{equation}
f\left( \bar{\psi},\psi \right) =M\left( \bar{\psi}\psi \right) ,
\label{PFD4.18y}
\end{equation}
where $M$ is a polynomial in $\bar{\psi}\psi $. With (\ref{PFD4.18x}) at
hand, the equation (\ref{PFD4.18}) becomes
\begin{equation}
\frac{1}{8}\left( \partial _{[\mu }\eta _{\nu ]}\right) \Pi ^{\mu \nu
}=\gamma \theta +\partial _{\mu }\rho ^{\mu },  \label{PFD4.18z}
\end{equation}
where
\begin{eqnarray}
&&\Pi ^{\mu \nu }\equiv \bar{\psi}\left[ \gamma ^{\mu },\gamma ^{\nu }\right]
\left( \partial _{\rho }g_{1}^{\rho }\right) -\left( \partial _{\rho
}g_{2}^{\rho }\right) \left[ \gamma ^{\mu },\gamma ^{\nu }\right] \psi
\nonumber \\
&&-4\left( \left( \partial _{\left. {}\right. }^{\left[ \mu \right. }\bar{%
\psi}\right) g_{1}^{\left. \nu \right] }-g_{2}^{\left[ \mu \right. }\left(
\partial _{\left. {}\right. }^{\left. \nu \right] }\psi \right) \right)
\nonumber \\
&&+\frac{\partial ^{R}\left( \left( \partial _{\rho }\bar{\psi}\right)
g_{1}^{\rho }+g_{2}^{\rho }\left( \partial _{\rho }\psi \right) \right) }{%
\partial \psi }\left[ \gamma ^{\mu },\gamma ^{\nu }\right] \psi  \nonumber \\
&&-\bar{\psi}\left[ \gamma ^{\mu },\gamma ^{\nu }\right] \frac{\partial
^{L}\left( \left( \partial _{\rho }\bar{\psi}\right) g_{1}^{\rho
}+g_{2}^{\rho }\left( \partial _{\rho }\psi \right) \right) }{\partial \bar{%
\psi}}.  \label{8}
\end{eqnarray}
The left-hand side of (\ref{PFD4.18z}) is $\gamma $-exact modulo $d$ if
there exists a real, bosonic function, involving only the undifferentiated
Dirac fields, $M^{\rho \mu \nu }$, which is antisymmetric in its last two
indices
\begin{equation}
M^{\rho \mu \nu }=-M^{\rho \nu \mu },  \label{7}
\end{equation}
such that
\begin{equation}
\Pi ^{\mu \nu }=\partial _{\rho }M^{\rho \mu \nu }.  \label{8x}
\end{equation}
From (\ref{8}), we observe that the existence of such functions $M^{\rho \mu
\nu }$ is controlled by the functions $g_{1}^{\mu }$ and $g_{2}^{\mu }$. The
most general form of the functions $g_{1}^{\mu }$ is
\begin{equation}
g_{1}^{\mu }=\psi g^{\mu }+\sum\limits_{n=1}^{4}\gamma ^{\nu _{1}}\cdots
\gamma ^{\nu _{n}}\psi g_{\nu _{1}\ldots \nu _{n}}^{\mu },  \label{9}
\end{equation}
where $g^{\mu }$ and $g_{\nu _{1}\ldots \nu _{n}}^{\mu }$ are some real,
bosonic functions in the undifferentiated Dirac fields. Now, from (\ref%
{PFD3.51}) it results that
\begin{equation}
g_{2}^{\mu }=g^{\mu }\bar{\psi}+\sum\limits_{n=1}^{4}g_{\nu _{1}\ldots \nu
_{n}}^{\mu }\bar{\psi}\gamma ^{\nu _{n}}\cdots \gamma ^{\nu _{1}}.
\label{10}
\end{equation}
Inserting (\ref{9})--(\ref{10}) in (\ref{8}), we arrive at
\begin{eqnarray}
&&\Pi ^{\mu \nu }=\left( \partial _{\rho }\left( \bar{\psi}\psi \right)
\right) \left( \frac{\partial ^{R}g^{\rho }}{\partial \psi }\left[ \gamma
^{\mu },\gamma ^{\nu }\right] \psi -\bar{\psi}\left[ \gamma ^{\mu },\gamma
^{\nu }\right] \frac{\partial ^{L}g^{\rho }}{\partial \bar{\psi}}-4\sigma
^{\rho [\mu }g^{\nu ]}\right)  \nonumber \\
&&+\partial _{\rho }\left( -\sum\limits_{n=1}^{4}g_{\nu _{1}\ldots \nu
_{n}}^{\rho }\left( \bar{\psi}\gamma ^{\nu _{n}}\cdots \gamma ^{\nu _{1}}%
\left[ \gamma ^{\mu },\gamma ^{\nu }\right] \psi -\bar{\psi}\gamma ^{\nu
_{1}}\cdots \gamma ^{\nu _{n}}\left[ \gamma ^{\mu },\gamma ^{\nu }\right]
\psi \right) \right)  \nonumber \\
&&+\sum\limits_{n=1}^{4}\left( \frac{\partial ^{R}g_{\nu _{1}\ldots \nu
_{n}}^{\rho }}{\partial \psi }\left[ \gamma ^{\mu },\gamma ^{\nu }\right]
\psi -\bar{\psi}\left[ \gamma ^{\mu },\gamma ^{\nu }\right] \frac{\partial
^{L}g_{\nu _{1}\ldots \nu _{n}}^{\rho }}{\partial \bar{\psi}}-4\sigma
_{\left. {}\right. }^{\rho [\mu }g_{\nu _{1}\ldots \nu _{n}}^{\nu ]}\right)
\times  \nonumber \\
&&\times \left( \left( \partial _{\rho }\bar{\psi}\right) \gamma ^{\nu
_{1}}\cdots \gamma ^{\nu _{n}}\psi +\bar{\psi}\gamma ^{\nu _{n}}\cdots
\gamma ^{\nu _{1}}\left( \partial _{\rho }\psi \right) \right)  \nonumber \\
&&+4\sum\limits_{n=1}^{4}g_{\nu _{1}\ldots \nu _{n}}^{\rho
}\sum\limits_{k=1}^{n}\left( \left( \partial _{\rho }\bar{\psi}\right)
\gamma ^{\nu _{1}}\cdots \gamma ^{\nu _{k-1}}\sigma ^{\nu _{k}[\mu }\gamma
^{\nu ]}\gamma ^{\nu _{k+1}}\cdots \gamma ^{\nu _{n}}\psi \right.  \nonumber
\\
&&\left. +\bar{\psi}\gamma ^{\nu _{n}}\cdots \gamma ^{\nu _{k+1}}\sigma
^{\nu _{k}[\mu }\gamma ^{\nu ]}\gamma ^{\nu _{k-1}}\cdots \gamma ^{\nu
_{1}}\left( \partial _{\rho }\psi \right) \right) .  \label{13ww}
\end{eqnarray}
The right-hand side of (\ref{13ww}) is of the form $\partial _{\rho }M^{\rho
\mu \nu }$ if
\begin{eqnarray}
&&\frac{\partial ^{R}g^{\rho }}{\partial \psi }\left[ \gamma ^{\mu },\gamma
^{\nu }\right] \psi -\bar{\psi}\left[ \gamma ^{\mu },\gamma ^{\nu }\right]
\frac{\partial ^{L}g^{\rho }}{\partial \bar{\psi}}-4\sigma ^{\rho [\mu
}g^{\nu ]}=C^{\rho \mu \nu },  \label{13mx} \\
&&n=1,\;g_{\nu _{1}}^{\rho }=\delta _{\nu _{1}}^{\rho }Q\left( \bar{\psi}%
\psi \right) ,  \label{14ww}
\end{eqnarray}
where $C^{\rho \mu \nu }=-C^{\rho \nu \mu }$ are some constants (and not
some $4\times 4$ matrices) and $Q$ is an arbitrary polynomial in $\bar{\psi}%
\psi $. Since in four spacetime dimensions there are no such constants, they
must vanish, which imply that
\begin{equation}
\frac{\partial ^{R}g^{\rho }}{\partial \psi }\left[ \gamma ^{\mu },\gamma
^{\nu }\right] \psi -\bar{\psi}\left[ \gamma ^{\mu },\gamma ^{\nu }\right]
\frac{\partial ^{L}g^{\rho }}{\partial \bar{\psi}}-4\sigma ^{\rho [\mu
}g^{\nu ]}=0.  \label{11ww}
\end{equation}
The solution to the last equation reads as
\begin{equation}
g^{\rho }=\bar{\psi}\gamma ^{\rho }\psi N\left( \bar{\psi}\psi \right) ,
\label{11}
\end{equation}
where $N$ is an arbitrary polynomial in $\bar{\psi}\psi $. In consequence, $%
\Pi ^{\mu \nu }$ expressed by (\ref{8}) can be written like in (\ref{8x}) if
the functions (\ref{9})--(\ref{10}) are of the form
\begin{eqnarray}
g_{1}^{\mu } &=&\psi \left( \bar{\psi}\gamma ^{\mu }\psi \right) N\left(
\bar{\psi}\psi \right) +\gamma ^{\mu }\psi Q\left( \bar{\psi}\psi \right) ,
\label{11x} \\
g_{2}^{\mu } &=&\left( \bar{\psi}\gamma ^{\mu }\psi \right) N\left( \bar{\psi%
}\psi \right) \bar{\psi}+\bar{\psi}\gamma ^{\mu }Q\left( \bar{\psi}\psi
\right) .  \label{11y}
\end{eqnarray}
By means of (\ref{11x})--(\ref{11y}) we deduce
\begin{eqnarray}
&&\left( \partial _{\mu }\bar{\psi}\right) g_{1}^{\mu }\left( \bar{\psi}%
,\psi \right) +g_{2}^{\mu }\left( \bar{\psi},\psi \right) \left( \partial
_{\mu }\psi \right) =\partial _{\mu }\left( \bar{\psi}\gamma ^{\mu }\psi
P\left( \bar{\psi}\psi \right) \right)  \nonumber \\
&&+s\left( \mathrm{i}\left( \psi ^{*}\psi -\bar{\psi}\bar{\psi}^{*}\right)
\left( P\left( \bar{\psi}\psi \right) -Q\left( \bar{\psi}\psi \right)
\right) \right) ,  \label{12ww}
\end{eqnarray}
where $P$ is a polynomial in $\bar{\psi}\psi $ defined by $N\left( \bar{\psi}%
\psi \right) =dP\left( \bar{\psi}\psi \right) /d\left( \bar{\psi}\psi
\right) $. The relation (\ref{12ww}) shows that the last two terms from (\ref%
{PFD3.50}) produce a trivial deformation, which can always be removed by
setting
\begin{equation}
g_{1}^{\mu }\left( \bar{\psi},\psi \right) =0,\;g_{2}^{\mu }\left( \bar{\psi}%
,\psi \right) =0.  \label{12x}
\end{equation}
Then, with the help of (\ref{PFD4.18y}) it follows that
\begin{equation}
a_{0}^{\left( \mathrm{Dirac}\right) }=M\left( \bar{\psi}\psi \right) .
\label{12y}
\end{equation}

Inserting (\ref{PFD4.18y}) and (\ref{12x}) in (\ref{PFD4.17}), we finally
find that the interacting lagrangian at order two in the coupling constant
takes the form
\begin{eqnarray}
b_{0}^{\left( \mathrm{int}\right) } &=&-\frac{\mathrm{i}}{4}\bar{\psi}\gamma
^{\mu }\left( \partial ^{\nu }\psi \right) \left( hh_{\mu \nu }-\frac{3}{2}%
h_{\mu \sigma }h_{\nu }^{\sigma }\right) -\frac{1}{4}\left( \bar{\psi}%
\mathrm{i}\gamma ^{\mu }\left( \partial _{\mu }\psi \right) -m\bar{\psi}\psi
\right) \times  \nonumber \\
&&\times \left( h_{\alpha \beta }h^{\alpha \beta }-\frac{1}{2}h^{2}\right) -%
\frac{\mathrm{i}}{32}\bar{\psi}\gamma ^{\mu }\left[ \gamma ^{\alpha },\gamma
^{\beta }\right] \psi \left( h\partial _{[\alpha }h_{\beta ]\mu }\right.
\nonumber \\
&&\left. -h_{\mu }^{\sigma }\partial _{[\alpha }h_{\beta ]\sigma }+h_{\alpha
}^{\sigma }\left( 2\partial _{[\beta }h_{\sigma ]\mu }+\partial _{\mu
}h_{\beta \sigma }\right) \right) +\frac{1}{2}M\left( \bar{\psi}\psi \right)
h.  \label{PFD4.19}
\end{eqnarray}
The formulas (\ref{PFD4.15x}), (\ref{PFD4.16}) and (\ref{PFD4.19}) reveal
the full, interacting, second-order deformation of the solution to the
master equation.

\end{document}